\documentclass[12pt,preprint]{aastex}
\usepackage{lscape}

\accepted{}
\journalid{}{}
\articleid{}{}

\shortauthors{Mohanty et al..}
\shorttitle{T Tauri Phase in Brown Dwarfs}

\received{2003 March}
\begin{document}

\def\lam{$\lambda$}
\def\tross{${\tau}_{R}$ }
\def\hal{{{\rm H}\alpha} }
\def\rc{R_C}
\def\ic{I_C}
\def\rcm{R_{Cm}}
\def\icm{I_{Cm}}
\def\jm{J_{m}}
\def\rco{R_{Co}}
\def\ico{I_{Co}}
\def\jo{J_{o}}
\def\av{A_V}
\def\ar{A_{R_C}}
\def\ai{A_{I_C}}
\def\exj{A_{J}}
\def\kr{k_{R_C}}
\def\ki{k_{I_C}}
\def\rad{{\mathcal{R}}_{\ast}}
\def\dist{\mathcal{D}}
\def\mass{{\mathcal{M}}_{\ast}}
\def\lum{{\mathcal{L}}_{bol}}
\def\logten{log_{10}}

\def\li{\ion{Li}{1}\ }
\def\na{\ion{Na}{1}\ }
\def\pot{\ion{K}{1}\ }
\def\oxy{\ion{O}{1}\ }
\def\hel{\ion{He}{1}\ }
\def\cal{\ion{Ca}{2}\ }
\def\nit{\ion{N}{2}\ }
\def\sul{\ion{S}{2}\ }

\def\kms{km s$^{-1}$\ }
\def\kmsp{km s$^{-1}$ pix$^{-1}$\ }
\def\cms{cm s$^{-1}$\ }
\def\cmc{cm$^{-3}$\ }
\def\cmss{cm$^{2}$ s$^{-1}$\ }
\def\cmcs{cm$^{3}$ s$^{-1}$\ }

\def\mdot{$\dot{M}$ }
\def\msun{M$_\odot$\ }
\def\rsun{R$_\odot$\ }
\def\lsun{L$_\odot$\ }
\def\mj{M$_J$\ }

\def\teff{T$_{e\! f\! f}$~}
\def\gv{{\it g}~}
\def\vsini{{\it v}~sin{\it i}~}
\def\vrad{v$_{\it rad}$~}
\def\lbol{L_{\it bol}}
\def\lhal{L_{H\alpha}}
\def\fhal{F_{H\alpha}}
\def\fcal{{\mathcal{F}}_{CaII}}
\def\fcont{{\mathcal{F}}_{cont}}
\def\fbol{F_{\it bol}}
\def\lx{L_{X} }
\def\eqwhal{EW_{H\alpha}}
\def\eqwcal{EW_{CaII}}
\def\alom{{\alpha}{\Omega} }
\def\ross{R_{0} }
\def\cots{{\tau}_{c} }
\def\fchal{{\mathcal{F}}_{c\hal} }
\def\h2o{H$_2$O}
\def\vcal{r_{CaII} }
\def\fex{{\mathcal{F}}_{excess}}

\title{The T Tauri Phase Down to Nearly Planetary Masses: \\Echelle Spectra of 82 Very Low-Mass Stars and Brown Dwarfs}

\author{Subhanjoy Mohanty\altaffilmark{1},  Ray Jayawardhana\altaffilmark{2}, Gibor Basri\altaffilmark{3}}

\altaffiltext{1}{Harvard-Smithsonian Center for Astrophysics, Cambridge, MA 02138.  smohanty@cfa.harvard.edu}
\altaffiltext{2}{Department of Astronomy \& Astrophysics, University of Toronto, Toronto, ON M5S 3H8, CANADA.  rayjay@astro.utoronto.ca}
\altaffiltext{3}{Astronomy Department, University of California, Berkeley, CA 94720.  basri@soleil.berkeley.edu}

\begin{abstract} 
Using the largest high-resolution spectroscopic sample to date of young, very low mass stars and brown dwarfs, we investigate disk accretion in objects ranging from just above the hydrogen-burning limit all the way to nearly planetary masses. Our 82 targets span spectral types from M5 to M9.5, or masses from 0.15 \msun down to about 15 Jupiters.  They are confirmed members of the $\rho$ Ophiuchus, Taurus, Chamaeleon I, IC 348, R Coronae Australis, Upper Scorpius and TW Hydrae star-forming regions and young clusters, with ages from $\lesssim$1 to $\sim$10 Myr.  The sample contains 41 brown dwarfs (spectral types $\geq$M6.5). We have previously presented high-resolution optical spectra for roughly half the sample; the rest are new.  This is a close to complete survey of all confirmed brown dwarfs known so far in the regions examined, except in $\rho$ Oph and IC 348 (where we are limited by a combination of extinction and distance).  We find that: {\it (1)} classical T Tauri-like disk-accretion persists in the sub-stellar domain down to nearly the deuterium-burning limit; {\it (2)} while an $\hal$ 10\% width $\gtrsim$ 200 \kms is our prime accretion diagnostic (following our previous work), permitted emission lines of \cal, \oxy and \hel are also good accretion indicators, just as in CTTs (we caution against a blind use of $\hal$ width alone, since inclination and rotation effects on the line are especially important at the low accretion rates in these objects); {\it (3)} the \cal 8662\AA~ line flux is an excellent {\it quantitative} measure of the accretion rate in very low-mass stars and brown dwarfs (as in higher-mass CTTs), correlating remarkably well with the \mdot obtained from veiling and $\hal$-modeling; {\it (4)} the accretion rate diminishes rapidly with mass -- our measurements support previous suggestions that $\dot{M}$ $\propto$ ${\mass}^2$ (albeit with considerable scatter), and extend this correlation to the entire range of sub-stellar masses; {\it (5)} the fraction of very low-mass stellar and brown dwarf accretors decreases substantially with age, as in higher-mass stars; {\it (6)} at any given age, the fraction of very low-mass stellar and substellar accretors is comparable to the accretor fraction in higher-mass stars; and {\it (7)} a number of our sources with infrared excesses arising from dusty disks do not evince measurable accretion signatures, with the incidence of such a mismatch increasing with age:  this implies that disks in the low mass regime can persist beyond the main accretion phase, and parallels the transition from the classical to post-T Tauri stage in more massive stars.  These strong similarities at young ages, between higher-mass stars and low-mass bodies close to and below the hydrogen-burning limit, are consistent with a common formation mechanism in the two mass regimes.  
\end{abstract}


\section{Introduction}
There has been much recent interest and activity in investigating the origin and early evolution of sub-stellar bodies. Some theorists, most recently Padoan \& Nordlund (2004), have suggested that objects all the way from solar-mass stars to brown dwarfs form the same way, via `turbulent fragmentation'. Others have proposed that brown dwarfs are in fact ``stellar embryos'', ejected from newborn multiple systems through dynamical gravitational interactions before accreting sufficient mass to become full-fledged stars (Reipurth \& Clarke 2001; Bate et al. 2003). A detailed comparison between the properties of brown dwarfs and stars, in their infancy, can help distinguish between these scenarios.  

Recent observations make a compelling case that objects near and below the sub-stellar boundary undergo a T Tauri phase, remarkably similar to their higher mass counterparts. For example, excess emission in the near- and mid-infrared, indicative of dusty disks, appears to be common in young brown dwarfs (e.g., Muench et al. 2001; Natta et al. 2002; Jayawardhana et al. 2003). So do spectroscopic signatures of disk-accretion, such as broad, asymmetric H$\alpha$ lines (Jayawardhana, Mohanty \& Basri 2002, 2003 -- hereafter JMB02, JMB03; Muzerolle et al. 2003 -- hereafter MHCBH03; Natta et al. 2004 -- hereafter N04). Irregular high-amplitude photometric variations, seen in some brown dwarfs, may also be related to accretion (Scholz \& Eisl{\"o}ffel 2004). What's more, a number of objects around the hydrogen burning limit exhibit forbidden line emission, usually associated with jets and winds in T Tauri and Class I sources (e.g., Fernandez \& Comeron 2001; Barrado y Navascues, Mohanty \& Jayawardhana 2004; MHCBH03; N04; Barrado y Navascues \& Jayawardhana 2004). 

In this paper, we greatly extend the investigation of accretion in the very low mass stellar and sub-stellar regimes, with the largest sample of high-resolution optical spectra for such objects to date.  Our sample includes more than 80 targets, extending from just above the hydrogen burning limit all the way down to masses approaching the planetary domain. Thus, we are able to explore accretion in very low mass stars and brown dwarfs over an order of magnitude in mass and several orders in derived accretion rates, look for correlations with parameters such as age and disk-related infrared excess, and compare these properties to those of higher mass T Tauri stars in order to probe the formation mechanism. 

Before proceeding further, it is important to emphasize two points that bear on the question of brown dwarf origins.  The first concerns the inherent uncertainty in certifying sub-stellar status in young objects.  In the field, the presence of Lithium is a litmus test for brown dwarfs, with objects evincing \li in their spectra bound to be less massive than $\sim$0.06 \msun, and thus indubitably sub-stellar.  The test fails at ages of a few Myrs, however, since even stars have not yet depleted their Lithium content over such short timescales.  In the absence of such a direct test, sub-stellarity in young objects is inferred via techniques that are highly dependent on theoretical models.  At present, the latter models are quite poorly anchored in the young very low-mass stellar and substellar regimes, given the paucity of empirically determined parameters (temperature, mass, radius) for such objects.  Consequently, the masses determined from these models are subject to significant uncertainties; combined with reasonable observational errors, this can lead to mass offsets of order 50\%.  Under the circumstances, at early ages, {\it no} current technique can distinguish between brown dwarfs and very low-mass stars that lie close the stellar/sub-stellar boundary: a young ``star'' with inferred mass 0.1 \msun may well be a 0.07 \msun brown dwarf instead, and vice versa.  For ease of nomenclature, one may certainly assume a working definition of sub-stellarity: e.g., within the context of the widely-used Lyon evolutionary models, a spectral type of $\sim$M6 is usually adopted as the dividing line between stars and brown dwarfs at a few Myrs (here we conservatively adopt M6.5; see further below).  It must be kept in mind, though, that any such boundary is only indicative; the true division may lie, for instance, anywhere from M5.5--6.5.  Debates over whether an individual young object, at say M6, is ``really'' a star or a brown dwarf are currently impossible to settle, and thus meaningless. 
 
Given this difficulty in distinguishing between the stellar and sub-stellar domains at early ages, how does one choose a sample to explore the formation mechanism of brown dwarfs?  The first solution is to include a large number of relatively late-type sources that are likely to be sub-stellar in spite of the above uncertainties: e.g., a few Myr-old M8 object, with inferred mass $\sim$0.03 \msun, is quite certain to be a brown dwarf even if its true mass is a factor of 2 higher.  Our present sample satisfies this criterion.  The second, more important answer is that a distinction between very low mass stars and brown dwarfs is not truly relevant to the formation question.  The `ejection' and `turbulent fragmentation' hypotheses seek to explain how very low mass objects can form out of molecular clouds, when the average thermal Jeans mass in the latter is usually of order a solar-mass.  This is a problem, however, that is not restricted solely to the formation of brown dwarfs, but extends to very low mass stars as well.  If we do not know the formation mechanism of a 0.08 \msun brown dwarf, we are also in the dark about the origins of a 0.1 \msun star.  For example, in the simulations by Bate et al. (2003), brown dwarfs {\it and} very low mass stars are produced through ejection.  In this context, the statement often made that brown dwarfs form (or do not form, depending on one's preferred hypothesis) like stars draws a spurious distinction at the hydrogen-burning limit: no formation scenario put forward so far depends physically on whether an object will eventually fuse hydrogen or not.  The real physical distinction lies between higher-mass solar-type stars, whose formation appears well addressed by the thermal Jeans mass formulation, and bodies -- both stars and brown dwarfs -- with masses that are lower by an order of magnitude (or more).  It is the formation of all the latter objects that is the fundamental question.  To probe this issue, it is acceptable to lump together young objects that are both close to and below the stellar/sub-stellar boundary, without having to precisely determine sub-stellar status; the crucial comparison is between this group and significantly higher-mass T Tauri stars, in order to explore whether the same formation process is at play in both mass regimes.   We carry out such comparisons in the present work.  

In view of the above discussion, we adopt the following nomenclature in this paper.  ``Very low mass stars'' (VLMS) are defined as objects with spectral types in the range M5 -- $<$M6.5.  At the 1--10 Myr ages we are concerned with here, this formally corresponds to masses, from Lyon theoretical tracks, of 0.15 $\gtrsim$ $\mass$ $>$ 0.075 \msun (Baraffe et al. 1998).  ``Brown dwarfs'' (BDs) are defined as young sources with spectral types $\geq$M6.5, corresponding to theoretical masses $\lesssim$ 0.075 \msun.  ``Low-mass objects'' will refer to the combined VLMS and BD sample, i.e., all young sources that are M5 or later ($\lesssim$ 0.15 \msun).  Finally, the latter sample will often be compared to ``higher-mass stars'', by which we mean young stars with spectral types $\sim$ K0 -- M4, corresponding to theoretical masses 2 $\gtrsim$ $\mass$ $\gtrsim$ 0.25 \msun at a few Myrs.  The low and high mass ranges we choose allow clear comparisons between stars close to a solar mass, and objects smaller than a solar mass by an order of magnitude or more.

\section{Sample and Observations}
Our 82 targets are located in the $\rho$ Oph, Taurus, IC348, Cha I, R Coronae Australis (R CrA), Upper Sco and TW Hydrae star-forming regions and young clusters.  These regions span ages from $\lesssim$1 Myr to $\sim$10 Myr; the objects range in mass from about 0.15 \msun down to $\sim$15 \mj (age and mass determinations discussed in \S3).  The sources are all {\it bona-fide} cluster members: they are selected from low-resolution spectroscopic studies which have already established their membership in the respective regions (Luhman \& Rieke 1999; Brice\~{n}o et al. 2002; Luhman et al. 2003a, b; Comeron, Neuhauser \& Kaas 2000; Fernandez \& Comeron 2001; Barrado y Navascu\'{e}s, Mohanty \& Jayawardhana 2004; Gizis 2002; Mohanty et al. 2004a; Mart\'{i}n, Delfosse \& Guieu 2004).  We further support youth and membership through the detection of lithium (\li \lam 6708; except in the coolest/faintest objects, which have very low S/N in the blue), radial velocities, and the presence of narrow atomic alkali lines (\na, \pot) indicative of low gravity.  We have previously presented high-resolution optical spectra for 33 of these targets -- 4 in $\rho$ Oph, 1 in R CrA, 3 in the TW Hydrae Association (TW HyA), 4 in Taurus, 10 in IC348 and 11 in Upper Sco (JMB02; JMB03; Mohanty, Jayawardhana \& Barrado y Navascu\'{e}s 2003; Barrado y Navascu\'{e}s, Mohanty \& Jayawardhana 2004; Mohanty et al. 2004a, b).  The remaining 49 are new, comprising 22 sources in Taurus, 5 in IC348, 13 in Cha I and 9 in Upper Sco.  

Our sample includes most known BDs (i.e., spectral type $\geq$ M6.5) in the surveyed regions, down to $\ic$ $\approx$ 19.  In Taurus, Cha I, Upper Sco, TW HyA and R CrA, this includes nearly all the sub-stellar objects identified to date (barring a total of $\sim$ 10 objects; of course, this is at the time of writing -- there are many ongoing surveys for new BDs).  In $\rho$ Oph and IC348, our high-resolution optical spectroscopy is limited by a combination of extinction (in both) and distance (in IC348).  Consequently, we have only 4 sources in $\rho$ Oph, including 2 of the 5 established BDs (Luhman \& Rieke 1999); similarly, we could observe only a fourth of the $\sim$20 BDs identified so far in IC348 (Luhman et al. 2003a).  Moreover, we are somewhat biased towards accretors in IC348, in the M5--$<$M6 range.  Luhman et al. (2003a) have found $\sim$50 objects at these types in IC348; with limited telescope time in hand, we preferentially observed those that were likely to be accreting (based on the $\hal$ equivalent widths in Luhman et al.'s published low-resolution spectra).  As a result, 4 out of our 6 IC348 stars in the M5--$<$M6 range indeed turn out to be accretors; in contrast, the majority of Luhman et al.'s objects at these spectral types actually evince low $\hal$ emission (equivalent widths $\lesssim$ 10\AA) and are unlikely to be accreting.  However, with our IC348 sample contributing a mere 6 VLMS in this spectral type range to begin with, this bias does not affect our analyses of accretor fractions (as we discuss later).  Finally, there is no such significant bias in our IC348 sample at types $\geq$ M6 (where we have observed two-thirds -- 9 out of 14 -- of the objects with $\ic$ $<$ 19 in Luhman et al.'s study, regardless of whether the latter authors' low-resolution spectra suggested possible accretion or not).  

We obtained high-dispersion optical spectra of the sample over numerous observing runs from May 2002 to December 2003.  For $\rho$ Oph, Taurus, IC348 and Upper Sco, we used the High Resolution Echelle Spectrograph (HIRES; Vogt et al. 1994) on the Keck I telescope at Mauna Kea.  With the a slit-width of 1''.15, the 2-pixel binned spectral resolution is $R \approx$ 33,000.  The instrument yielded 15 spectral orders in the 6390--8700\AA~ wavelength region, with gaps between the orders, covering a variety of features related to youth and accretion.  The data were reduced in standard fashion with IDL routines, as described in Basri et al. (2000).  For the Cha I, R CrA and TW HyA southern targets, we used the Magellan Inamori Kyocera Echelle spectrograph (MIKE; Bernstein et al. 2002) on the Baade 6.5m telescope at Las Campanas Observatory, Chile.  The instrument provided simultaneous spectra over $\sim$3200--4800\AA~ in the blue and $\sim$4800--8800\AA~ in the red, with overlapping orders in each wavelength range.  Most spectra were obtained with a slit-width of 1'', giving resolutions of $R \approx$ 25,000 and 19,000 in the blue and red respectively.  A few spectra in TW HyA were obtained with a narrower, 0''.7 slit; these observations, and more details on the Magellan spectra in general, are given in Mohanty, Jayawardhana \& Barrado y Navascu\'{e}s (2003).  The data reduction was analogous to that employed for the Keck spectra.  Finally, we point out that all spectra shown in this paper have been smoothed by a 3-pixel boxcar.      

\section{Adopted Parameters}
Our analysis of accretion, in subsequent sections, requires estimates for various physical quantities, such as effective temperature (\teff), mass and line flux.  The values we have adopted for the pertinent parameters, and the rationale for our choices, are described below.

{\it Spectral Type}: Spectral types for our $\rho$ Oph, Taurus, IC348 and ChaI targets -- i.e., for most of our sample -- are from Luhman \& Rieke 1999, Brice\~{n}o et al. 2002, Luhman et al. 2003a and 2003b, and Luhman 2004.  All these works employ the same optical low-resolution SED-fitting technique, with combined dwarf and giant standards as templates, to infer the spectral types, ensuring that the latter are all mutually consistent (for $\rho$ Oph, the types are derived in the near-IR, but subsequently pegged to the optical standards; Luhman \& Rieke 1999).  For LS-RCrA-1, our only target in R CrA, we adopt a spectral type of M6.5, as found by Barrado y Navascu\'{e}s, Mohanty and Jayawardhana (2004) from optical medium- and high-resolution spectra.  This is consistent with the $\sim$M6 determined for us by K. Luhman, with the low-resolution SED-fitting technique (K. Luhman, pvt. comm., 2004).  

For targets in Upper Sco and TW HyA, our adopted types are from Ardila, Mart\'{i}n \& Basri 2000, Mart\'{i}n, Delfosse \& Geieu 2004 (for the USco DENIS targets), and Gizis 2002.  These authors use various optical molecular band indices for spectral typing, not quite the same methodology as SED-fitting.  Now, we primarily use spectral types in this paper for estimating various  physical parameters, such as \teff and mass, for our accreting objects.  For a self-consistent analysis, therefore, we must ensure that the spectral types from band indices for our two accretors in Upper Sco and TW HyA -- USco DENIS 160603 and 2MASS 1207-3932 (\S5 and Table 1) -- are consistent with the types deduced from SED-fitting in all our other accretors.  We have checked this by comparing our high-resolution optical spectra of these two objects to the spectra of targets with types derived from SED-fitting.  Specifically, we have examined the TiO bandheads around 8440\AA~, which are highly sensitive to \teff (Mohanty et al. 2004a) and should therefore reflect spectral type variations.  We find that the spectra of USco DENIS 160603 and 2MASS 1207-3932 are consistent with SED-fitting types to within $\pm$0.5 subclasses. For instance, 2MASS 1207-3932 has a spectral type of M8 from band indices (Gizis 2002); we find that its TiO band shape and strength is comparable to that of targets with spectral type M8$\pm$0.5, where the latter are derived from SED-fitting.  Since the intrinsic uncertainty in the SED method is also of order $\pm$0.5 subclasses, we may adopt the quoted types for USco 160603 and 2MASS 1207-3932 unchanged.  Similar analysis of all the {\it non}-accreting Upper Sco and TW HyA sources shows that their spectral types, from band-indices, agree to within $\pm$0.5--1 subclass with the SED-fitting types.  The latter variations are somewhat larger than ideally desirable; nevertheless, since we do not use spectral types for any detailed analysis of the non-accretors, we may adopt their quoted values unchanged as well, without affecting the results of this paper\footnote{For two sources in our sample -- USco 130 and 131 -- Ardila, Mart\'{i}n \& Basri assign spectral types of M7--M8 and M6.5 respectively, from $I$-$J$ color alone and not low-resolution spectroscopy.  Bases on our high-resolution TiO analysis, we assign both a spectral type of M7, consistent with their color-based types.}.  

{\it Effective Temperature}:  We estimate \teff from spectral types via the conversion scale specified in Luhman et al. (2003a).  This scale is explicitly constructed to agree with the predictions of the Lyon evolutionary tracks (Baraffe et al. 1998, Chabrier et al. 2000), which are the most widely used tracks today in the low-mass regime.  This does not guarantee the absolute accuracy of the inferred \teff: their veracity depends on the validity of the Lyon models, which remain somewhat uncertain for very young low-mass objects (Mohanty et al. 2004a,b).  However, the numbers are likely to be in the right ballpark (within 100--200K; Mohanty et al. 2004a), as are the {\it relative} differences in \teff over the M5--M9.5 spectral subclasses considered here (Mohanty et al., in prep.).  Since we are primarily interested in the relative variations and trends within our sample, the Luhman-scale is adequate here.

{\it Age}: For the star-forming regions and young clusters in our sample, we adopt ages from the literature derived via comparisons to theoretical H-R diagrams.  The ages are: $\lesssim$1 Myr for $\rho$ Oph (Luhman \& Rieke 1999); $\sim$1.5 Myr for Taurus\footnote{Brice\~{n}o et al. 2002 state 1--2 Myr; we therefore adopt a mean age of 1.5 Myr.  This agrees with the $\sim$1.5 Myr age found by White et al. 1999 for the GG Tau quadruple system in Taurus.  Luhman et al. 2003a compare their IC348 sample to that of Brince\~{n}o et al.'s low-mass sample in Taurus, and show that the Taurus sources appear slightly younger overall than the IC348 ones: they ascribe a median age of $\sim$1 Myr to Taurus and $\sim$2 Myr to IC348.  Our adopted ages for Taurus and IC348 -- 1.5 and 2 Myr respectively -- are commensurate with the statements by Brice\~{n}o et al., Luhman et al. and White et al., and preserve the slight age difference between the two regions found by Luhman et al.} (Brice\~{n}o et al. 2002; Luhman et al. 2003a); $\sim$2 Myr for IC348 and ChaI (Luhman et al. 2003a, Luhman 2004); $\sim$3 Myr for R CrA (Wilking et al. 1997); $\sim$5 Myr for Upper Sco (Preibisch et al. 2002); and $\sim$10 Myr for TW HyA (Webb et al. 1999; Jayawardhana et al. 1999).  For Taurus, IC348, ChaI and Upper Sco, these mean ages are inferred from the Lyon tracks, while D'Antonella \& Mazzitelli models (1994, 1997; hereafter DM models) have been used for $\rho$ Oph, R CrA and TW HyA.  Ideally, all the ages should be based on the same set of tracks.  However, this does not affect our statistical analysis of accretion versus age (\S 5.4), in which we ignore R CrA and TW HyA altogether (since we have only 1 accretor in each).  We do include $\rho$ Oph; however, a visual inspection of the Taurus and $\rho$ Oph HR diagrams (Brice\~{n}o et al. 2002; Luhman \& Rieke 1999) suggests that $\rho$ Oph is indeed somewhat younger than Taurus in an absolute sense: sources with similar \teff are on average more luminous in $\rho$ Oph. Thus, with Taurus at about 1.5 Myr on the Lyon tracks, $\lesssim$1 Myr appears valid for $\rho$ Oph even with the latter models (similarly, though we ignore TW HyA in our age analysis, we note that it is surely older than the other regions, regardless of the precise age implied by a set of tracks).
 
{\it Mass and Radius}:  A unique mass can be determined for any source from the theoretical evolutionary tracks, given either \teff and bolometric luminosity ($\lbol$), or \teff and age.  The masses of all our IC348 and Taurus accretors (except CHFT-BD-Tau 4) have been supplied by K. Luhman (pvt. comm. 2004), and inferred from [\teff,$\lbol$] comparisons to the Lyon theoretical models (the track comparisons are shown in Brice\~{n}o et al. 2002 and Luhman et al. 2003a, though the masses are not explicitly stated there).  The \teff used here are from spectral types, as described earlier, and $\lbol$ derived by combining $J$ magnitudes, extinction estimates and bolometric corrections (Brice\~{n}o et al. 2002).  For the remaining accretors, we have obtained masses by comparing [\teff, age] to the Lyon models, where `age' is the mean age of the cluster to which the source belongs (given above for the various regions).

In practice, $\lbol$ or age has a minimal effect on the inferred mass:  our cool sources evolve down approximately vertical Hayashi tracks at nearly constant \teff for the first $\sim$10 Myr, so \teff alone is sufficient to derive a relatively precise mass; the $\lbol$ or age provides only a small correction.  This is why we are justified above in using the mean cluster age as a proxy for the real age of a given source, even though objects in any specified region scatter around this mean.  This is also why it is not inconsistent to use Lyon tracks to derive masses for accretors in $\rho$ Oph, Upper Sco and TW HyA, even though the mean ages of these regions are from DM models: while it would be technically more consistent to use mean cluster ages also from the Lyon models in these cases, any model-dependent age variations are too small to affect the derived masses at these ages once \teff is specified (and our \teff in all cases {\it are} ultimately from the Lyon models, which the Luhman spectral type--\teff scale is based on).  As an example of how little mass is affected by the adopted age, we note that for an object with \teff = 2700K (M8 on the Luhman scale), the Lyon models imply a mass of 30 \mj at 1 Myr and 35 \mj at 10 Myr -- a negligible difference for our purposes here.  It should be pointed out that age does make a difference, in the Lyon tracks, for the very lowest masses ($<$ 20 \mj), which get rapidly cooler and fainter from 1 to 10 Myr.  However, we only use masses for accretors: among the latter, our two sources at $\lesssim$20 \mj (KPNO-Tau 4 and 12) have masses derived via K. Luhman's explicit comparison of [\teff, $\lbol$] to the Lyon tracks, without making any simplying assumptions about the constancy of \teff with age.    

The difference between the Lyon and DM models {\it is} important when it comes to the mass that corresponds to a given \teff.  At young ages, and for cool objects like ours, the DM models tend to yield masses that are about a factor of 1.5--2 lower than the Lyon ones, for a specified temperature (MHCBH03).  Insofar as the lowest masses covered in our study therefore, our estimates using the Lyon models are more conservative than studies employing DM tracks (e.g., MHCBC03, N04).  Even with Lyon tracks, however, we go down to $\sim$15 \mj, i.e., to  nearly the planetary-mass (deuterium-burning) limit at 12 \mj.  

Finally with regard to mass, we note that in the Lyon models, the stellar/sub-stellar boundary at 0.075 \msun occurs at \teff $\approx$ 3000--2930K, over the $\lesssim$1--10 Myr range considered here.  This corresponds to M6--M6.5 on the Luhman spectral type--\teff conversion scale, and accounts for our conservative working definition of BDs (\S 1) as objects at types $\geq$ M6.5.  While this is accurate for later types, however, one should always keep in mind (as stated in \S 1) that sources close to the boundary could fall either way, due to both spectral typing and evolutionary model uncertainties : an object that is M6.5 in our scheme, and whose mass is determined via theoretical (Lyon) tracks, can either be a VLMS or a BD, though it is guaranteed to lie close to the sub-stellar edge.  

Lastly, accretion rates for some of the higher mass, classical T Tauri stars (CTTs) we shall examine have been derived by other groups using veiling measurements, and masses and radii from DM models (\S5.3).  Since \mdot from veiling depends on $\rad/\mass$, we must modify these \mdot to Lyon masses and radii, in order to use them in our study.  These sources are all in Taurus; we have re-derived their masses and radii from Lyon models through the procedures outlined above (i.e., converting spectral type to \teff, and then inferring mass and radius by comparing \teff to the Lyon tracks for the adopted mean Taurus age of 1.5 Myr).  Now, radius depends more sensitively on age than mass does. However, since our 1.5-Myr estimate for Taurus is derived from the same Lyon tracks, not DM ones, our analysis is self-consistent insofar as the theoretical models used.  Moreover, the spread in Taurus ages is $\sim$ $\pm$2 Myr; over this time and for CTTs masses, the Lyon tracks indicate a variation in $\rad$ (and hence in \mdot for a given mass and veiling) of only $\sim$20\%.  This uncertainty is negligible, since the \mdot from veiling are uncertain by factors of $\sim$3--5 to begin with (e.g., Gullbring et al. 1998).
  
{\it Equivalent Width}: We measure emission equivalent widths for $\hal$, \hel \lam6678, \oxy \lam8446 and \cal \lam8662.  Now, in cool mid- to late-M objects such as our targets, no `true' continuum exists in the optical, only a {\it pseudo-continuum} comprising a plethora of overlapping molecular lines.  All our widths are measured relative to the latter, and are thus more precisely pseudo-equivalent widths; we drop the `pseudo-' appellation henceforth only for the sake of concision.  Furthermore, a perusal of the \hel \lam6678 region in field (Main Sequence, non-accreting) mid- to late-M dwarfs reveals a bump at the line position, even in objects with fairly weak activity.  Whether this is true chromospheric emission in \hel, or merely a bump coincidentally produced by molecular absorption on either side of the line position, is unclear.  We have therefore compared the spectra of all our young objects (M5--M9.5) to that of the field dwarf Gl 406 (M6), and calculated the \hel \lam6678 equivalent width after {\it excluding} the bump at this position in the Gl 406 spectrum (this spectral region changes very little, in field dwarfs from $\sim$M5--M9, so using a single M6 dwarf template is justified for the level of accuracy desired here).  While this may underestimate the widths slightly, it precludes our assuming \hel emission where in fact there is none; we err on the side of caution.  Similarly, the \oxy \lam8446 line sits on top of a TiO absorption band (bandheads at \lam\lam\lam8432, 8442, 8451).  We have thus calculated \oxy widths relative to the pseudo-continuum formed by this band, through comparison with young, clearly non-accreting objects of the same spectral type in our sample with no evidence of \oxy emission (field dwarfs are not appropriate templates here, since the TiO band strengths, though primarily dependent on \teff, do change in going from very young objects at a given spectral type, to field dwarfs of the same type but at much higher gravity).  Within the context of this methodology, the errors in our inferred equivalent widths are about $\pm$10\% for widths $\gtrsim$ 1\AA, and $\pm$0.1\AA~ for widths $<$ 1\AA.  Our detection limit is $\sim$ 0.1\AA.  
  
{\it Continuum Flux}: Our derivation of accretion rates (\S5.2, 5.3) involves calculating \cal emission fluxes ($\fcal$).  To infer the latter from the observed line equivalent widths ($\eqwcal$), we must know the photospheric continuum flux ($\fcont$) underlying the \cal lines.  In previous studies of CTTs, $\fcont$ has been estimated from the observed $I$-magnitude.  However, we shall examine masses all the way from CTTs to BDs, spanning $\Delta$\teff $>$1000K.  Over this range in temperatures, changes in the broad-band $I$-magnitude reflect a plethora of variations in photospheric opacity not necessarily confined to the immediate spectral vicinity of \cal.  To circumvent this problem, we use the $\fcont$ predicted by the latest Lyon synthetic spectra for a specified \teff and surface gravity.  We use the Lyon STAR-DUSTY 1999 and 2000 models\footnote{Openly available on the internet at ftp://ftp.ens-lyon.fr/pub/users/CRAL/fallard} (Allard, Hauschildt \& Schwenke 2000, hereafter AHS00; Allard et al. 2001), \teff inferred from our spectral types, and a surface gravity of log[g]=4.0 (cgs units).  

The Lyon DUSTY models incorporate the formation of atmospheric dust grains through chemical equilibrium calculations, and include both the resulting depletion of grain-forming species and dust opacity.  These models are appropriate to the mid- to late-M spectral types our low-mass sample covers, since field dwarf studies indicate grain formation becomes important by late M\footnote{In practice, dust affects the models, at the wavelengths of the \cal IRT, only at the lowest \teff (latest types) in our sample: over most of our \teff range, grains have a negligible effect on the \cal region in the synthetic spectra.  This agrees with the observation that in field dwarfs, dust appears in blue spectral regions at a type of $\sim$M7 and in the red regions appropriate to \cal at M8--M9, and that in very young objects dust appears at even later types than in field dwarfs (see detailed discussion in Mohanty et al. 2004a).}.  They are also appropriate for hotter, higher mass stars, in which no dust forms: since dust formation in the models is handled self-consistently, grains simply do not appear in the DUSTY models at high enough \teff.  The DUSTY models come in two flavors: BD-DUSTY and STAR-DUSTY.  Both use the same set of TiO opacities in the optical (AMES-TiO line-lists), but different sets of \h2o opacities in the near-IR (AMES-\h2o line-lists in the BD models and MT-\h2o in the STAR; see AHS00 and references therein).  Neither flavor is preferred over the other: both BD-DUSTY and STAR-DUSTY have some inadequacies in reproducing the observed near-IR SED of M stars, but both match the optical colors and SEDs of M stars quite well (AHS00).  Under the circumstances, we choose the STAR-DUSTY models, for the following reason.  In our analysis of accretion rate versus \cal flux (\S 5.2), we will compare our low-mass, mid- to late-M sample to hotter, higher-mass CTTs (mostly K7--M0, but two at K3--K5).  Only the STAR-DUSTY models are currently available over a large enough range of \teff (2500--5000K) to simultaneously analyse both the high and low-mass regimes.  In particular, STAR-DUSTY 1999 models are available up to 5000K; STAR-DUSTY 2000 models, which incorporate the latest improvements in dust opacity and are therefore better for the mid- to late M's where dust might be important, are available up to 4000K.  We therefore use the 2000 models for all sources with \teff $\leq$ 4000K (K7 and later types), and the 1999 ones for the two CTTs with \teff $>$4000K.  

Our choice of surface gravity is dictated by the mean value indicated by the Lyon evolutionary tracks, for low-mass objects that are a few Myrs old.  In practice, the continuum flux depends predominantly on \teff, and only negligibly on gravity as long as log[g] $\approx$ 4.0$\pm$0.5 dex (well within the limits predicted by the Lyon models for ages $\sim$1--10 Myrs).  The $\fcont$ adopted is the average value in the synthetic spectra over 8600--8700\AA~, which covers the  8662\AA~ \cal component of interest here (\S 5).  This technique has been profitably used previously for calculating $\hal$ fluxes in field M dwarfs, by Mohanty \& Basri (2003); we refer the interested reader to that work for further details.  Once the continuum flux $\fcont$ is known, the line flux is calculated as: $\fcal$ = $\fcont$ $\times$ $\eqwcal$ $\times$ $(1 + \vcal)$.  Here $\vcal$ $\equiv$ $\fex/\fcont$ is the accretion-induced veiling near the \cal lines, i.e., a measure of the excess continuum emission ($\fex$) produced by the accretion shock relative to the photospheric continuum.  For objects from other studies, we adopt the quoted $\vcal$ (\S5.2); for our own sample, we estimate $\vcal$ $\approx$ 0 in most cases (\S 5.3).  

There might certainly be systematic offsets between the fluxes we infer in this manner, and the true values, due to uncertainties in the synthetic continuum fluxes.  However, this effect is likely to be small since, as we will show, our values in the CTT regime agree quite well with those derived by other researchers using $I$-magnitudes.  Moreover, we will mainly be concerned in this paper with the correlation between \cal fluxes and accretion rates.  Once we quantitatively establish this relationship for objects with known \mdot and \cal fluxes found via our technique, we will use it to infer accretion rates for other objects with \cal flux derived in the {\it same} fashion.  Since the relationship is anchored using \mdot that are calculated independent of the \cal fluxes, the ``true'' value of the flux is not key to the analysis; it is only important that all the fluxes be derived via a {\it consistent} methodology.  

{\it Rotation Velocity}: We derive rotation velocities (\vsini) of our targets by cross-correlating with a ``spun-up'' template of a slowly rotating standard.  Multiple spectral orders ($\sim$ 6), selected on the basis of an absence of strong telluric features, strong gravity-sensitive features and stellar emission lines, were used for the analysis.  Following White \& Basri (2003), we used a combination of giant and dwarf spectra for our templates.  A detailed discussion of the cross-correlation technique is given in Mohanty \& Basri (2003).  Our errors are, conservatively, $\sim$ $\pm$2 \kms for \vsini $\lesssim$ 20 \kms, $\pm$2.5--3 \kms for \vsini of 20--50 \kms, and $\pm$5 \kms for \vsini of 50--70 \kms ($\sim$60\kms being the highest \vsini in our sample).

\section{Accretion Diagnostics}
It is useful to begin with a summary of the expected behaviour, in the presence of accretion, of the main diagnostic spectral lines included in our spectra: $\hal$ (6563\AA), \hel (6678\AA), \oxy (8446\AA) and \cal (8662\AA, the reddest component of the \cal infrared triplet).

\subsection{$\hal$}
In classical T Tauri stars (CTTs), the widely accepted paradigm of magnetospheric accretion holds that stellar field lines threading the disk channel material from the inner edge of the disk onto the stellar surface.  The gas in these `accretion columns' falls in at nearly free-fall velocities, resulting in an accretion-shock as it strikes the surface; $\hal$ is produced both in the shock region and in the infalling gas.  Consequently, $\hal$ emission is generally much stronger in CTTs than in non-accreting weak-line T Tauri stars (WTTs), where $\hal$ arises only from chromospheric activity.  The $\hal$ line-profiles in CTTs are also distinctive, evincing broad line-wings extending out to a few hundred kilometers per second from line-center; the broadening is due to both the high (free-fall) velocity of the accreting gas and the large damping wings associated with Stark-broadening in this very optically thick line (Muzerolle, Calvet \& Hartmann, 2001).  Often, the CTTs $\hal$ emission is also asymmetric, due to inclination effects (e.g., the star/disk may occlude part of the infall) and/or absorption by an accompanying outflowing wind (e.g., Alencar \& Basri 2000).  

The strength of the $\hal$ line has long been used as a simple basis for distinguishing between CTTs and WTTs, with objects exhibiting $\hal$ equivalent widths $>$ 10\AA~ classified as CTTs.  However, this diagnostic is not applicable over any large range of spectral types, due to the `contrast effect'.  Basically, as one moves to cooler (later type, lower mass) objects, the same $\hal$ flux stands out better against the increasingly faint photosphere.  As a result, the equivalent width of even chromospheric $\hal$ can exceed 10\AA~ at late spectral types.  As an alternative, White \& Basri (2003; hereafter WB03) have proposed a quantitative discriminant between CTTs and WTTs based on $\hal$ line {\it profiles}.  In CTTs, the accretion shock produces excess continuum emission in the optical that fills in underlying photospheric absorption lines; such `veiling' is generally a sure sign of accretion.  WB03 point out that, empirically, the full-width of the $\hal$ line at 10\% of the line-peak invariably exceeds 270 \kms in veiled CTTs, regardless of spectral type.  Thus an $\hal$ 10\% width $\gtrsim$ 270 \kms can by itself be adopted as a good accretion indicator, without a laborious check for veiling.

Subsequently, JMB03 concluded that WB03's accretion criterion, based primarily on CTTs observations and linked explicitly to the presence of veiling, is too restrictive when applied to the VLMS and BD regimes.  Accretion rates in the latter ($\sim$ 10$^{-10}$--10$^{-12}$ {\msun}yr$^{-1}$) are generally 1--3 orders of magnitude lower than in CTTs, too low for any appreciable veiling (Muzerolle et al. 2000, 2003).  At such puny rates, Stark-broadening of the $\hal$ line is also negligible; the line-broadening is dominated by the infall velocities alone, and thus smaller than in CTTs.  Furthermore, VLMS and BDs are less massive than CTTs, implying lower free-fall velocities for the gas and thus a further reduction in the extent of the $\hal$ line-wings.  From an examination of young low-mass objects in IC348, JMB03 showed that other accretion signatures, such as strong emission in the permitted lines of \hel, \oxy and \cal and asymmetries in the $\hal$ line profile, are associated with $\hal$ 10\% widths as low as 200 \kms.  They proposed, therefore, that an $\hal$ 10\% width $\gtrsim$ 200 \kms is a better accretion diagnostic for VLMS and BDs.
  
JMB03's suggested accretion cutoff is also supported by recent infrared spectroscopy.  N04 show that young VLMS and BDs with $\hal$ 10\% width $>$ 200 \kms generally also exhibit Pa$\beta$ emission associated with accretion.  Moreover, they find that in objects above this cutoff, the $\hal$ 10\% width correlates very well with the mass accretion rate derived through other means.  In the present paper, therefore, we adopt the 200 \kms $\hal$ criterion as our main accretion diagnostic (subject to the caveats discussed below).  This will be supported by line-profile asymmetries, as well as emission in other accretion-related lines (\S4.2) where applicable.  

Given that the infall rates in young VLMS and BDs are generally very small, and that the $\hal$ line-wings are likely to be dominated by infall velocities, three caveats must be borne in mind when using the $\hal$ line profile as an accretion diagnostic.  First, the rotation velocity of the object becomes important.  As we will show, the intrinsic full-width of the $\hal$ line in chromospherically active, non-accreting young VLMS and BDs in our sample is usually $\sim$ 100--150 \kms.  In a VLMS or BD with \vsini~ $\gtrsim$50 \kms, the full-width of the $\hal$ profile will be rotationally broadened by a further $\gtrsim$100 \kms.  Thus, with an adopted accretion cutoff of 200 \kms, a merely active but rapidly rotating low-mass object can appear to be accreting.  This effect is indeed present in a few of our targets, as we will discuss in \S 5.1.  

Second, since it is not the absolute infall velocities but rather the apparent ones along the line of sight that produce the $\hal$ line-wings, inclination effects are important - more so than in CTTs, where inclination-independent Stark-broadening usually dominates in the wings.  In an accretor oriented pole-on to our line of sight, most of the high-velocity infalling gas moves along the plane of the sky; hence the emission is concentrated at line-center (e.g., Muzerolle et al. 2003; hereafter MHCBH03).  Thus, in a weakly accreting VLMS or BD seen pole-on, the $\hal$ profile will be narrow and symmetric, possibly with a 10\% width $<$ 200 \kms, reminiscent of chromospheric activity instead of accretion.  This effect too is apparent in our sample, and must be corrected for by examining other accretion-related lines (\S 5.1).   

Finally, the observed $\hal$ 10\% width depends on the line optical depth.  At very low accretion rates, there simply may not be enough infalling gas to push the width above our 200 \kms cutoff.  An examination of the $\hal$ models profiles and comparisons to observations shown in MHCBH03 indicates that this effect is likely to be important for mass infall rates $\lesssim$ 10$^{-12}$ \msun{yr$^{-1}$}; of course, it is exacerbated by the inclination-angle effect discussed above.  In summary, adopting an $\hal$ 10\% width $\gtrsim$ 200 \kms as an accretion indicator is probably not sufficient to identify the weakest accretors (\mdot $\lesssim$ 10$^{-12}$ \msun{yr$^{-1}$}).  However, it is good for a {\it conservative} estimate of the number of accretors in our sample, especially when combined with the other accretion diagnostics addressed in the next section.
  
Here we would like to clarify the physical basis for JMB03's 200 \kms accretion cutoff for VLMS and BDs. For masses $\lesssim$ 0.1\msun, free-fall velocities from {\it infinity} are of order 200 \kms, which would simplistically lead to $\hal$ full-widths of $\sim$400 \kms.  However, the gas is expected to flow in from the inner edge of the disk; a variety of studies indicate that this lies at $\sim$ 2--3 sub-stellar radii in the VLMS and BD regime (e.g., Mohanty et al. 2004c).  Free-fall velocities from these radii are only $\sim$150 \kms; since the most likely inclination angle for a random distribution of orientations is $\sim$60$^{\circ}$, the apparent infall velocity is further reduced on average to $\sim$125 \kms, implying an $\hal$ full-width of $\sim$250 \kms.  Moreover, since we consider the full-width at 10\% of the peak, and not at the true `base' of the line, we measure less than the full range of infall velocities (e.g., the width at say 5\% is larger; 10\% is an ad hoc value chosen to ensure a good measurement of the width without confusion by the photospheric continuum).  Finally, since the line-width depends on the optical depth, which decreases as one moves to larger offsets from the mean velocity, we will always measure a lower limit to the true range of gas velocities (an effect that is exacerbated with declining accretion rate).  These considerations indicate that a lower limit of 200 \kms in $\hal$ 10\% width is a physically reasonable accretion criterion.  As JMB03, MHCBH03 and N04 have shown, other accretion indicators are indeed present in objects with $\hal$ 10\% widths larger than 200 \kms, so this limit is empirically justified as well.  

\subsection{\hel, \oxy and \cal}
Permitted emission lines of \hel, \oxy and \cal are also associated with accretion in CTTs (Muzerolle, Hartmann \& Calvet 1998, hereafter MHC98) -- in particular, \hel \lam\lam5876 and 6678, \oxy \lam\lam7773 and 8446, and the \cal infrared triplet (IRT) lines \lam\lam\lam 8498, 8542 and 8662.  Of these, our spectra include \hel \lam6678, \oxy \lam8446 and \cal \lam8662.  In their detailed study of CTTs, MHC98 showed that: the \oxy emission, as well as the broad (FWHM $\gtrsim$100 \kms) components of the \cal IRT and \hel lines, probably originate in the magnetospheric infall region; \oxy \lam8446 appears in emission only in CTTs with accretion rates (\mdot) $\gtrsim$ 10$^{-8}$ \msun{yr$^{-1}$}; and \hel, \oxy and \cal line strengths all increase with \mdot in CTTs, with the \cal IRT exhibiting the most striking correlation.  In young VLMS and BDs, the recent analyses by JMB03 and MHCMH03 reveal the following.  

{\it \oxy}: JMB03 find \oxy \lam8446 in some, though not all, of their low-mass IC348 accretors (defined as objects with $\hal$ 10\% width $\geq$ 200 \kms); the emission profile is broad, with FWHM $\gtrsim$ 100 \kms.  None of their non-accretors shows this line in emission.  MHCBH03 do not report on the behavior of permitted \oxy.  However, by modeling the $\hal$ line profile, they have derived infall rates for objects in their sample, including some of the accretors in which JMB03 detect \oxy \lam8446.  The inferred \mdot are of order 10$^{-10}$ \msun{yr$^{-1}$}, showing that in VLMS and BDs, this line can appear in emission at much lower accretion rates than in CTTs.  This is not surprising: a weak emission feature, that would be invisible against the bright photospheres of CTTs, is expected to show up better against the intrinsically faint photospheres of much cooler low-mass objects (the `contrast effect' mentioned earlier).  

{\it \hel}: MHCBH03 find that emission in \hel \lam5876 is ubiquitous, in both accreting as well as chromospherically active VLMS and BDs.  They do not discuss the behavior of \hel \lam6678 in their sample.  JMB03 do find \hel \lam6678 emission in one IC348 accretor (IC348-165), and do not report any emission in their non-accretors\footnote{A review of the spectra we presented in that paper shows \hel \lam6678 emission is also present in IC348-265, as noted in Table 1.  With an $\hal$ 10\% width slightly less than 200 \kms, this was not classified as an accretor by JMB03; however, we will argue in \S5 that this object is in fact likely to be accreting as well.}.  

{\it \cal}: MHCBH03 find widespread emission in the \cal IRT as well.  A closer perusal of their results shows that the \lam\lam8498 and 8542 IRT components are present in both accreting and merely active objects; however, the \lam8662 component occurs almost exclusively in accreting objects (9 of their 11 \lam8662 detections come from accretors, though not all their accretors exhibit this line).  Similarly, JMB03 detect \cal \lam8662 in some but not all of their IC348 accretors; in two of their three detections, the emission is relatively broad ($\gtrsim$ 100 \kms)\footnote{MHCBH03 state that broad emission in \oxy, \hel and \cal is present only in the earliest type accretor in their sample, Lk$\hal$ 358 (K7--M0), and not in any of their later-type accreting objects.  This is a little surprising at first, since JMB03 point out broad emission in some of the same low-mass objects in MHCBH03's sample.  The problem lies in the definition of ``broad'': MHCBH03 define this as a FWHM $\gtrsim$ 200 \kms, while JMB03, and we in this paper, define it as a FWHM $\gtrsim$ 100 \kms, in order to remain consistent with the definition in MHC98.  With this defining value, it is likely that \oxy, \hel and \cal emission from some of MHCBH03's accretors is broad as well, at least in targets in common with JMB03's.}.  Crucially, and in agreement with MHCBH03, they do not find any \cal \lam8662 emission in the non-accretors.   

Thus, emission in \oxy \lam8446, \hel \lam6678 and \cal \lam8662 seems reasonably well-correlated with accretion in VLMS and BDs, as in CTTs: while not detected in all the low-mass accretors, these lines appear preferentially associated with accretion, and not activity.  This assertion seems particularly safe to make for \cal \lam8662, for which the combined size of the JMB03 and MHCBH03 samples is large.  We conclude that emission in \hel \lam6678, \oxy \lam8446 and \cal \lam8662 is suggestive of accretion, and certifies accretion when allied with an $\hal$ 10\% width $\gtrsim$ 200 \kms.  We will re-examine this issue with our data in \S5.  In contrast, based on MHCBH03's results, \hel \lam5876 and \cal \lam\lam8498 and 8542 are not good accretion indicators in the low-mass regime, appearing in emission in both accreting and active objects.  We will not discuss the latter lines any further.   

\section{Results and Discussion}

\subsection{Observed Accretion and Ouflow}
Our accretion and activity results are summarized in Table 1.  $\hal$ profiles for half the sample have been shown in our previous work (JMB02; JMB03; Mohanty, Jayawardhana \& Barrado y Navascu\'{e}s 2003; Barrado y Navascu\'{e}s, Mohanty \& Jayawardhana 2004); profiles for the new objects in this paper are plotted in Figs. 1 and 2 (all accretors, old and new, in Fig. 1; new active objects in Fig. 2).  The majority of our 82 targets evince narrow and symmetric $\hal$ lines, with 10\% widths $\sim$ 100--150 \kms (Fig. 2).  Such emission is expected from chromospheric activity alone, so we classify these objects as non-accretors.  19 objects in our sample, however, have $\hal$ 10\% widths $>$ 200 \kms (Fig. 1), as well as low rotation velocities (\vsini $<$ 20 \kms).  Given the small implied rotational broadening, and the much narrower profiles in the sources that are merely active, the very broad emission in these 19 targets is highly unlikely to result from a combination of activity and rotation.  Instead, such profiles are expected for gas accreting at nearly free-fall velocities (\S4.1); we thus classify these sources as ``bona-fide'' accretors.  Asymmetries in many of their $\hal$ profiles, analogous to those often observed in CTTs, support this conclusion.  

A further two objects, IC348 355 (spectral type M8) and USco 75 (M6), also have $\hal$ 10\% widths slightly above 200 \kms; however, they are very rapid rotators, with a \vsini of 45 and 63 \kms respectively.  Their large $\hal$ 10\% widths are thus likely to result primarily from rotational broadening, not accretion.  This conclusion is strengthened by their $\hal$ equivalent widths (which are unaffected by \vsini effects): these are relatively low, and comparable to those in other non-accretors of the same spectral type.  They also do not evince any emission in \hel, \oxy or \cal, unlike many of the accretors (see further below). We thus classify these two objects as {\it non}-accretors\footnote{Note that JMB03 classified IC348-355 as an accretor; however, they did not consider rotation effects.}.  

Finally, we note that while MHO-5 shows no signs of accretion in our spectrum -- its $\hal$ 10\% width is less than 200 \kms, and it has no emission in permitted \hel, \oxy or \cal -- MHCBH03 have detected strong forbidden [\oxy] emission at \lam\lam 6300 and 6363 from this object.  Since such emission is associated with mass outflows in CTTs, which in turn provide indirect evidence of ongoing accretion, MHCBH03 classify MHO-5 as an accretor, though its $\hal$ 10\% width is less than 200 \kms in their data too.  Our Keck spectrum of MHO-5 does not cover these forbidden \oxy lines, so we cannot confirm MHCBH03's result; under the circumstances, we simply adopt their evaluation of this object as a bona-fide accretor.  In this case, its relatively narrow $\hal$ profile is probably due to a pole-on viewing orientation, as MHCBH03 also point out.  

Now let us examine the prevalence of \hel, \oxy and \cal emission.  Out of the 15 targets in which \hel \lam6678 is detected, 11 are likely accretors by our $\hal$ analysis above.  Similarly, 13 of the 17 \cal \lam8662 detections are in accretors, as are all 12 of the \oxy \lam8446 detections (\cal emission shown in Fig. 3).  This strongly validates our previous conclusion, based on the results of JMB03 and MHCBH03, that emission in these lines correlates with ongoing accretion.  In turn, this result prompts us to re-examine the status of the few objects which have $\hal$ 10\% widths $<$ 200 \kms, but nevertheless do evince emission in at least one of these lines: GY 310, KPNO-Tau 4, KPNO-Tau 11, CFHT-BD-Tau 4, MHO-4, IC 348 256 and USco 128.
  
Table 1 shows that GY 310 and IC 348 256 in fact have $\hal$ 10\% widths reasonably close to our accretion cutoff of 200 \kms -- 144 and 180 \kms respectively, comparable to the 166 \kms in the apparent accretor MHO-5 discussed above.  Fig. 1 moreover reveals that the $\hal$ profiles of the two sources are somewhat asymmetric.  Finally, there is direct evidence for inner disks in both, from mid-IR excesses (Mohanty et al. 2004; Jayawardhana et al. 2003).  While the presence of a disk does not certify accretion, since circumstellar material may exist beyond the main accretion phase (\S5.5), it is certainly consistent with ongoing disk-accretion.  These factors, combined with emission in either \cal or \hel, leads us to classify GY310 and IC348 256 as ``probable'' accretors.   

Similarly, the $\hal$ 10\% widths in KPNO-Tau 4 and 11 and CFHT-BD-Tau 4 are not much less than 200 \kms (155, 175 and 182 \kms respectively).  However, their $\hal$ profiles appear quite symmetric (Fig. 1), similar to those in chromospherically active objects.  We therefore classify them as ``possible'' accretors, with the following caveats.  We have already seen that \cal \lam8662 emission is associated only with accreting objects in the rest of our sample, implying that any purely chromospheric component of this line is very small.  Nevertheless, even such a weak component may stand out in emission against the faint ultra-cool photosphere of KPNO-Tau 4 (`contrast effect'); at M9.5, this is the coolest object in our sample.  Consequently, a purely chromospheric origin for its \cal emission is not wholly implausible.  On the other hand, Mohanty \& Basri (2003) have shown that such activity falls off sharply, at least in field dwarfs, at $\sim$M9, with little or no $\hal$ emission at later types; Mohanty et al. (2002) have further argued that this fall-off is due to the increasing neutrality of the cool atmospheres with later type.  Whether the same applies to young objects, i.e., whether a very cool young BD like KPNO-Tau 4 can in fact sustain sufficient activity to produce the observed $\hal$ and \cal emission, is unclear.  Spectroscopy of a large sample of young ultra-cool objects (very few known so far) is needed to settle this question.  

In CFHT-BD-Tau 4, which at M7 is somewhat hotter than KPNO-Tau 4, the contrast effect is not as strong.  Nevertheless, it is still quite cool, and its \cal emission is also very weak, so chromospheric \cal cannot be completely ruled out.  It is known to have a disk from mid-IR and sub-mm/mm observations; however, the disk also appears relatively evolved (Sterzik et al. 2004).  These facts do not permit a clear distinction between activity and accretion, so `possibly accreting' is the most conservative conclusion.  Finally, in KPNO-Tau 11, the contrast effect is much less important in both above sources, since the latter is siginficantly hotter at M5.5.  However, KPNO-Tau 11 does not show any \cal \lam8662, only \hel \lam6678.  Though this line is also predominantly associated with accretion in our sample, we do find it, without any accompanying \cal \lam8662 emission, in two objects we classify as non-accretors (see below).  Thus KPNO-Tau 11 may not be accreting, only much more active than most of our sample.  Further observations are required to clarify its status.    

Lastly, MHO-4 and USco 128, while exhibiting clear \hel \lam6678 emission, evince $\hal$ 10\% widths significantly lower than 200 \kms (115 and 111 \kms), as well as highly symmetric $\hal$ profiles (Fig. 1).  We thus classify them as non-accretors; note that MHCBH03 have independently found these two sources to be non-accretors as well.  Presumably, their \hel \lam6678 emission arises from very strong chromospheric activity, though we caution that accretion cannot be completely ruled out: very low infall rates combined with severe inclination effects may conceivably produce the narrow and symmetric $\hal$ profiles we observe.  

Two further sources deserve mention: KPNO-Tau 14 in Taurus, and DENIS 161929 in Upper Sco.  While deriving \vsini (\S 3), we discovered that their cross-correlation functions are slightly double-peaked, a telltale signature of spectroscopic binaries (SB2s).  We are carrying out follow-up observations to confirm this finding; for the moment, we refrain from citing a \vsini for them, and classify them as possible SB2s.  Both have $\hal$ 10\% widths less than 200\kms, and no emission in \hel, \oxy or \cal.  A close look at their $\hal$ profiles does reveal some asymmetry; e.g., the profile of DENIS 161929 (Fig. 2) is similar to that of the ``probable'' accretors GY310 and IC348 256 (Fig. 1).  However, such asymmetries can also result from overlapping $\hal$ profiles in an SB2.  Without more definitive accretion signatures (broad $\hal$, or emission in other accretion-related lines as in GY310 and 1C348 256), we conservatively classify them as non-accretors for now.  As an aside, we note that KPNO-Tau 14 has a spectral type of M6, and DENIS 161929 of M8; if their SB2 status is confirmed, they would be the two latest-type (lowest mass) spectroscopic binaries discovered to date at ages of a few Myrs.  As such, they might prove very useful for testing theoretical evolutionary tracks for very young VLMS and BDs, through empirical mass and radius determinations. 
  
In summary, a total of 25 VLMS and BDs in our sample evince accretion signatures of some sort or another (broad $\hal$ and/or emission in other accretion-related lines).  Of these, 20 are ``bona-fide'' accretors: 19 based on $\hal$ profiles, and 1 (MHO-5) based on its forbidden [\oxy] emission in MHCBH03's data.  The remaining 5 sources (GY 310, IC348 256, KPNO-Tau 4, KPNO-Tau 11 and CFHT-BD-Tau 4) are classified as ``probable'' or ``possible'' accretors; further observations are required to verify their accretor status. 

Recently, Barrado y Navascu\'{e}s \& Mart\'{i}n (2003; hereafter BM03) have suggested equivalent width criteria for discriminating between infall and activity in both higher-mass stars as well as in VLMS and BDs.  They set the upper limit for chromospheric $\hal$ flux at log[$\lhal/\lbol$] = -3.3, the average saturation value observed in young open clusters.  This {\it flux} limit is then converted into a quantitative equivalent width cutoff for different spectral types.  Objects with $\hal$ widths above the cutoff appropriate to their spectral type are assumed to be accreting, while those below the limit are likely to be merely active.  Note that since the same $\hal$ flux becomes more prominent against a cooler photosphere (contrast effect), the limiting equivalent width derived by BM03 increases with later spectral type.

The BM03 scheme is specifically constructed for low-resolution spectra, which generally yield larger equivalent widths than high-resolution ones.  Later in the paper (\S 5.4), we will use previously obtained low-resolution data for our low-mass sample to derive accretor fractions based on the BM03 criteria.  It is useful to briefly discuss here, however, the results we will obtain.  In general, we find that the BM03 scheme performs reasonably well: targets that we find to be accreting are largely classified as such by BM03's technique, and similarly for non-accretors.  Nevertheless, the BM03 scheme systematically gives larger accretor fractions for our low-mass sample, when applied to low-resolution $\hal$ widths, compared to the fractions suggested by our detailed high-resolution analysis above.  There are two reasons for this.  First, sources that are more active than the norm for their spectral type, or are possibly chromospherically flaring, will be mis-classified as accretors by the BM03 criteria; their non-accretor status can only be deciphered from a high-resolution examination of their $\hal$ profiles such as undertaken here.  Conversely, weak accretors seen pole-on can have narrow and symmetric $\hal$ profiles at high-resolution, while their $\hal$ equivalent widths and fluxes (unaffected by inclination) can still be higher than that in active objects.  Such objects would appear to be merely active from our high-resolution analysis, while they would correctly be classified as accretors by BM03.  MHO-5 is a good example of this.  Its $\hal$ equivalent width at low resolution is substantially higher than in chromospherically active sources at the same spectral type (M6), implying ongoing accretion by the BM03 criteria.  However, the object appears a non-accretor in our high-resolution data -- its $\hal$ profile is narrow and symmetric, and it lacks other accretion-related lines such as \cal.  We are aware of its true accretor status only serendipitously, from the [\oxy] emission spotted by MHCBH03; other MHO-5 analogs may well lack such emission, and thus simply appear active in our high-resolution study.  In summary, our high-resolution analysis is more conservative in identifying accretion; while we may classify some accretors as active, we are unlikely to deduce accretion in merely active sources.  While the BM03 scheme is very useful for statistical analyses of large samples, we prefer our conservative approach for the classification of individual sources.

Accretion rates are generally much lower in the low-mass regime than in CTTs (see \S5.2 and 5.3 below); much of the preceding discussion underscores the difficulties that occasionally arise in distinguishing between infall and activity under these circumstances.  Nevertheless, out of the 25 sources in which we have identified some accretion signatures, 20 can be clearly classified as bona-fide accretors.  The latter sample includes: {\it (1)} an accretor at every spectral subclass from M5 to M9, {\it (2)} the lowest mass accreting BD known to date -- the Taurus M9 object KPNO-Tau 12 (mass $\sim$ 20 Jupiters), {\it (3)} the only known BD accretor in the 5 Myr-old Upper Sco region, where most stars have also stopped accreting -- DENIS 160603 (M7.5, mass $\sim$ 40 Jupiters), and {\it (4)} the oldest accreting BD currently known -- the 10 Myr-old TW HyA member 2MASS 1207-3932 (M8, mass $\sim$ 35 Jupiters).  Our observations of this last object have been discussed in detail in Mohanty, Jayawardhana \& Barrado y Navascu\'{e}s 2003.  Our diagnosis of accretion in that paper, based on $\hal$ and \hel emission, has since been buttressed by X-ray analyses (Gizis 2004), and the detection of accretion shock-induced UV emission (Gizis 2004, pvt. comm.).  We note that this BD is now also known to possess a surrounding disk, from mid-IR excess measurements (Sterzik et al. 2004). 
 
Next, we turn our attention to mass outflows, which are often associated with accretion in CCTS, and are widely identified in the latter through forbidden line emission.  We have already mentioned that MHO-5 shows strong [\oxy] emission in MHCBH03's data, though this region is not covered in our Keck spectrum.  We do not see any other optical forbidden lines (e.g., [\nit], [\sul]) in this object.  Similarly, Natta et al. 2003 see hints of [\nit] and [\sul] emission in Cha $\hal$ 2 and 6 (which we classify as accretors based on $\hal$), and in Cha $\hal$ 3 (classified here as a non-accretor); we do not see these lines in our spectra of these three objects (possibly due to lower S/N).  However, we do find very strong emission in [\oxy], [\nit] and [\sul] in our Magellan spectra of the accretor LS-RCrA-1, as discussed in Barrado y Navascu\'{e}s, Mohanty \& Jayawardhana 2004 (see also Fernandez \& Comeron 2001).  At a spectral type of $\sim$M6.5, this object resides at the stellar/sub-stellar boundary, and to date has offered the best evidence for the existence of outflowing jets and winds in the low-mass regime.  We have now detected, for the first time, weak [\oxy] emission in 2MASS 1207-3932 as well (Fig. 4).  At an age of 10 Myr, and a spectral type of M8 (implying a mass $\sim$ 35 Jupiters), this TW HyA member is the oldest, lowest mass BD known to evince any mass outflow signature (as well as the oldest known BD with accretion signatures, as discussed above).  The overall paucity of jet/wind detections in VLMS and BDs so far, compared to the CTT regime, is not surprising: outflow rates are expected to scale with infall rates, which are very small in the low-mass domain.  Nonetheless, our detections, combined with those by others, imply that such outflows can exist in accreting VLMS and BDs.

Taken together, the above results suggest that: {\it (1)} CTT-like accretion extends over the {\it entire} sub-stellar regime, down to the lowest mass BDs known, {\it (2)} outflows can also be present in accreting VLMS and BDs, as in CTTs, and {\it (3)} the accretion timescales in VLMS and BDs are comparable to those in higher-mass stars (as discussed in detail in \S5.4).  

\subsection{Accretion Rates}
In CTTs, accretion rates are usually determined through veiling measurements.  Unfortunately, infall rates in VLMS and BDs are usually far too small to produce any discernible veiling (WB03, MHCBH03).  For these objects, \mdot may be estimated through detailed modeling of the $\hal$ line profile (MHCBH03).  This process is laborious and time-consuming, however, and a more straightforward technique is highly desirable.  A possible solution is to employ \cal \lam8662 fluxes: as discussed earlier, \cal IRT component fluxes correlate very well with \mdot at least in higher-mass CTTs (MHC98).  In our low-mass sample, we have already shown that the presence of emission in the \lam8662 component is closely allied to ongoing accretion.  We now demonstrate that the flux in this line also correlates quantitatively with \mdot for these low masses, just as in CTTs.  For low-mass accretors with \cal emission, this will permit a direct derivation of \mdot without resorting to $\hal$ profile-modeling.  

MHC98, WB03 and MHCBH03 provide \cal \lam8662 equivalent widths, veiling and \mdot measurements for a number of objects, from CTTs to BDs.  The accretion rates are derived from either the observed veiling (all the CTTs) or detailed $\hal$ line-profile modeling (most of the VLMS and BDs).  We use this sample to probe the relationship between \cal flux and \mdot from the stellar to sub-stellar regimes.  Our analysis technique is as follows.   

(1) We convert the \cal equivalent widths quoted by MHC98 and MHCBH03 to fluxes through the technique described in \S3.  In the case of CIDA-1 and GM Tau, WB03 provide \mdot and veiling but not the \cal equivalent widths; however, we have access to the spectra they used, and derived the \cal widths in these ourselves.  (2) The veiling \mdot we use are from MHC98 and WB03; these are all calculated using the methodology outlined by Gullbring et al. 1998.  This is explicitly true of the WB03 values (derived from $\rc$-band veiling), and of most of the MHC98 values, taken directly from Gullbring et al.'s work (who use $U$-band veiling).  For a few heavily-veiled stars not analysed by Gullbring et al., MHC98 use \mdot from Hartigan, Edwards \& Ghandour 1995 (calculated from $V$-band veiling), whose assumptions leads to systematically higher \mdot than the Gullbring et al. technique; however, MHC98 have adjusted these to the Gullbring et al. scale as well, through multiplication by an appropriate factor.  (3) All these \mdot inferred from veiling are proportional to the assumed $\rad/\mass$, so we must ensure that the masses and radii used are consistent with the Lyon tracks adopted in this paper.  WB03 use Lyon tracks, so we adopt their \mdot unchanged; the \mdot in MHC98 are all based on DM tracks, so we have modified these \mdot to Lyon masses and radii (methodology given in \S 3).  (4) The \mdot from $\hal$ profile-modeling, on the other hand, are quite insensitive to mass and radius assumptions; thus, we adopt unchanged the \mdot found via such modeling by MHCBH03.  (5) We have left out unresolved binaries included in the MHC98 and WB03 studies, since binarity can vitiate the calculation of \mdot from veiling (Gullbring et al. 1998). 

The sources we examine, and their final adopted parameters, are listed in Table 2.  The resulting relationship between \cal \lam8662 flux and accretion rate is shown in Fig. 5a.  The two quantities clearly correlate remarkably well, over more than 4 dex in both \mdot and line flux.  MHC98 found such a relationship in CTTs, which we recover in Fig. 5a for \mdot between $\sim$10$^{-6}$ and 10$^{-8.5}$ \msun yr$^{-1}$.   In particular, Fig. 9 in MHC98 shows the correlation between their \mdot and \cal $\lambda$8542 fluxes, where the line equivalent widths are converted to fluxes using $I$-band magnitudes and veiling estimates.  We have used their veiling measurements, as well as their quoted $\lambda$8662 widths, which are very similar to their $\lambda$8542 ones; thus, any differences between their Fig. 9 and our Fig. 5a for the same objects are ascribable to our use of synthetic continuum fluxes to infer line fluxes, versus their use of $I$-magnitudes for the same purpose.  A comparison of the plots shows that our results are in fact very similar; while our derived \cal fluxes are systematically slightly higher than theirs, the trend in flux versus \mdot is identical (note that a few outliers in their plot are not included by us, since they are binaries).  This bolsters our confidence in the technique we have used to derive the line fluxes.  Compared to MHC98, our crucial new result is that the correlation extends all the way to VLMS and BDs, with accretion rates several orders of magnitude lower than in CTTs (most objects plotted with \mdot $<$ 10$^{-8.5}$ \msun yr$^{-1}$ are VLMS or BDs; see Table 2).  A linear fit to the data yields: 
$$\quad\qquad log\,(\dot{\rm M}) \,\, = \,\, 1.06 \,\, log\,(\fcal) \,\,-\,\,15.40 \qquad\qquad .......{\rm CTTs+VLMS+BDs}\eqno[1]$$  
While a correlation between \mdot and \cal flux is obvious, however, Fig. 5a also reveals significant scatter around our derived fit: the 1$\sigma$ scatter in \mdot around the fit is a factor of 3, and the largest deviations are up to factor of 6.  Now, emission lines and veiling in CTTs are known to vary over time, and variability is also observed in low-mass accretors (see discussion in N04).  However, it is unlikely that the observed scatter is due to such variablity combined with a difference in epoch between the \cal flux and \mdot measurements.  For all the low-mass objects (GM Tau, CIDA-1 and all IC348 sources), the \cal fluxes and \mdot have been derived from the {\it same} spectra.  Similarly, while spectra from different epochs have been used for the higher-mass CTTs (\cal measurements by MHC98, \mdot from Gullbring et al. 1998 and Hartigan, Edwards \& Ghandour 1995), MHC98 demonstrate that there is no significant change in the sources over this time (veilings found by MHC98 correlate very well with the veiling-dependent \mdot derived in the earlier studies).  In fact, a closer perusal of Fig. 5a shows that the scatter is systematic: all the low-mass objects lie somewhat above our fit, while most of the higher-mass CTTs fall somewhat below.  We illustrate this in Fig. 5b, by obtaining separate fits to the high- and low-mass regimes; the linear fits derived are:    
$$log\,(\dot{\rm M}) \,\, = \,\, 0.71 \,\, log\,(\fcal) \,\,-\,\,12.66 \qquad\qquad .......{\rm CTTs}\,\,\,\,\,\qquad\quad\eqno[2]$$  
$$log\,(\dot{\rm M}) \,\, = \,\, 0.93 \,\, log\,(\fcal) \,\,-\,\,15.03 \qquad\qquad .........{\rm VLMS+BDs}\eqno[3]$$  
We immediately see that the scatter around the individual fits is now much reduced: the 1$\sigma$ scatter in \mdot is a factor of $\sim$1.5 for each fit, and the largest deviations are at most a factor of $\sim$3.  The fits suggest an offset in the \cal flux--\mdot relationship between the high and low-mass regimes, with the low-mass objects evincing more \cal flux at a given \mdot.  

The reasons for this displacement, however, are unclear.  Systematic problems in the synthetic continuum fluxes, in going from the hotter CTTs to the cool low-mass sample, do not see responsible: the same offset appears even if we use observed $I$-band magnitudes to calculate the \cal fluxes of the high- and low-mass samples (not plotted).  Overestimation of the \cal flux in the low-mass accretors, through contamination by chromospheric \cal, is also unlikely: as discussed earlier, clearly {\it non}-accreting low-mass objects in our study simply do not show any \cal emission, so the purely chromospheric contribution is likely to be negligible in the low-mass accretors.  One might postulate, instead, that there are differences between the high- and low-mass regimes in the surface covering fraction of the accretion hot shock: if the shock covers relatively more surface area in the low-mass sources for a given \mdot, then their accretion-related \cal flux might also be higher than in CTTs, as observed.  However, this is not supported by the data: using the observed veilings, estimating the underlying photospheric fluxes (e.g., from synthetic spectra), and assuming a shock \teff $\sim$ 10$^4$K (e.g., MHCBH03), one easily finds that the covering fraction in the low-mass sources is $\ll$1\%, which is in fact quite similar to that in CTTs (as also shown by MHCBH03).  

One possibility is a mismatch between the \mdot from the ultraviolet $U$-band veiling (most of the higher-mass CTTs) and \mdot from optical diagnostics (all the low-mass objects: \mdot from optical veiling in GM Tau and CIDA-1, and from modeling optical $\hal$ profiles in the rest).  Another possibility is that the \cal flux in the high- and low-mass samples, while ultimately related to accretion in both cases, is not probing the same physical region in the two mass regimes.  Among the CTTs sample we have used, most of the sources (except the weakest accretors) evince broad \cal emission likely arising in the magnetospheric infall (MHC98).  Among the low-mass objects, such broad emission is seen only in GM Tau and CIDA-1; in the rest, the emission is narrow, and probably originates primarily in the accretion shock region on the (sub-)stellar surface.  If so, then differences in the \cal flux--\mdot correlation might be expected between the two mass regimes.  

Further exploration of this issue is beyond the scope of this paper, but important to pursue in the future.  For now, without a better idea of the underlying reasons for the shift between the low- and high-mass objects, we prefer the better fit given by equation [3] for the low-mass sample.  However, we will conservatively assume a factor of $\sim$5 uncertainty in the \mdot inferred from this fit: this amply covers both the small observed scatter in \mdot around the fit, as well as the systematic differences in \mdot implied by equations [1] and [3] (as we will see in \S 5.3 (Table 3), the \mdot derived from the two equations diverge by at most a factor of $\sim$3 for the \cal fluxes observed in our low-mass accretors; this arises mainly from the $\sim$0.4 dex difference in the intercept of the two equations, and slightly from the very small difference in their slopes).  Similarly, for the higher-mass CTTs, we prefer the fit given by equation [2] over that in equation [1].  However, we do not actually use \cal fluxes to calculate \mdot for CTTs in this paper; all the CTTs accretion rates in our subsequent analysis are adopted from the literature.  It is worth noting that these CTTs rates, derived from either veiling measurements or $\hal$ modeling, also have uncertainties of a factor of 3--5 (e.g., Gullbring et al. 1998; MHCBH03), comparable to the errors we adopt for our low-mass sample.  

The strong correlation between \cal flux and \mdot in low-mass accretors enables us to calculate their accretion rates directly from their observed \cal emission.  We accomplish this in the next section.  It is important to remember, however, that we do not see \cal in all our accreting VLMS and BDs, only in the majority.  Thus, while this technique for estimating \mdot allows us to avoid laborious $\hal$-modeling in many cases, such modeling is still required in very weak accretors where \cal emission is absent.  In particular, we are only sensitive to line equivalent widths $\gtrsim$ 0.1\AA.  Converting this to a a flux limit (different for various spectral types), and using equation [3], yields a cutoff in measurable \mdot of $\sim$ 10$^{-10.7}$--10$^{-11.7}$ \msun{yr$^{-1}$}, going from a spectral type of M5 to M9.5.  For still lower \mdot, $\hal$ modeling is necessary (though its usefulness runs out as well, below about 10$^{-12}$ \msun{yr$^{-1}$}).  

\subsection{Accretion and (Sub-)Stellar Mass}
Using the above relationships, we can now calculate accretion rates for all objects in our low-mass sample with measured \cal \lam8662 emission.  The calculation is rendered even simpler than in CTTs, because the small accretion rates in VLMS and BDs leads to minimal veiling ($\vcal$), as mentioned earlier, especially in the red part of the spectrum where \cal occurs (WB03).  Thus we can usually assume $\fcal$ = $\fcont$ $\times$ $\eqwcal$, dropping the veiling term that appears in the analogous equation in \S3.  Of course, the absence of veiling around \cal should be checked for: we have done this for our accretors with \cal emission, by comparing their high-resolution spectra to those of non-accreting targets with similar spectral types, and indeed find negligible veiling in most cases ($\vcal$ $\lesssim$0.1; in IC348 382, MHCBH03 find $\vcal$ = 0.2 whereas we see none, possibly because MHCBH03 assume an intrinsic spectral type of M6.5, while we use the most recently determined type of M5.5).  Veiling is present, though small, in only one of our sources -- LS-RCrA-1 ($\sim$0.15) -- as found (in the same spectrum we use) by Barrado y Navascu\'{e}s, Mohanty \& Jayawardhana 2004.

The accretion rates we infer from our observed \cal fluxes in the low-mass objects, using both equation [1] (simultaneous fit to the CTTs and low-mass sample) and equation [3] (separate fit to the low-mass sources), are cited in Table 3\footnote{Note that for 2 objects -- GM Tau and CIDA-1 -- our \cal fluxes have been combined with the \mdot derived by WB03 to {\it derive} the \mdot--\cal relationships in the last section.  We cannot therefore use these relationships to calculate new \mdot for them from the same \cal fluxes; for these two sources, we adopt the \mdot derived by WB03 (cited in both Tables 2 and 3).  For one further object -- IC34 336 -- we do not see any \cal emission, or any veiling; however, its TiO bands do not agree very well with those of other sources with similar spectral types.  This points to a possible small spectral typing uncertainty (of $\sim$ 1 spectral subtype; our type is from Luhman 2003a); under the circumstances, we do not derive an \mdot upper limit from the \cal non-detection, but adopt instead the \mdot found MHCBH03 from $\hal$ modeling (cited in Table 3).}.  We derive \mdot for both the ``bona-fide'' as well as ``probable'' and ``possible'' accretors; for accretors without discernible \cal emission, we quote \mdot upper limits based on a \cal equivalent-width detection-limit of 0.1\AA.  As advertised in the last section, the \mdot from equations [1] and [3] vary by at most a factor of $\sim$3 over our low-mass sample; our factor of $\sim$5 adopted uncertainty in \mdot covers both sets of values (though we prefer those from equation [3]).  Independent of which fit is used to derive \mdot, it is clear that accretion rates in our low-mass targets are generally far lower than in higher-mass CTTs.  To quantify this fall-off in \mdot with mass over the largest possible range in both parameters, we pool together 5 samples: {\it (i)} the objects used to derive the $\fcal$--\mdot relationship of equations [1]--[3], listed in Table 2; {\it (ii)} the accretors in our sample with \mdot deduced from equation [3], listed in Table 3; {\it (iii)} a few low-mass accretors from MHCBH03 not observed by us; {\it (iv)} additional CTTs from Gullbring et al. 1998, with \mdot from $U$-band veiling; and {\it (v)} additional CTTs from White \& Ghez 2001 (hereafter WG01), with \mdot from UV excess.  Objects in samples {\it (iii)}--{\it (v)} are all listed in Table 4.  

Some details about this combined sample are worth pointing out before proceeding further.  (1) There are a few low-mass objects (IC 348 165, 205, 382 and 415), with \mdot and \cal measurements by MHCBH03, that we have independently observed as well.  For these, we have used the MHCBH03 values in our derivation of eqn.[1,3], since MHCBH03 derive \mdot independently of \cal fluxes.  However, we have then applied eqn.[3] to {\it our} \cal measurements to derive new \mdot; for these objects, it is these new \mdot (cited in Table 3, not very different from those of MHCBH03 in Table 2) that we use in our \mdot--mass analysis.  Conversely, for MHO-5 and IC348 336, which do not show any \cal emission in our spectra but do have \mdot derived from $\hal$ profiles by MHCBH03, we use their values.  (2) All masses are from the Lyon tracks.  WG01 already cite Lyon $\mass$, so these are adopted unchanged; MHC98, Gullbring et al. 1998, and MHCBH03 cite $\mass$ from DM tracks, so these are modified to Lyon values (technique described in \S 3).  (3) All adopted \mdot are consistent with Lyon tracks.  \mdot from veiling are proportional to $\mass$/$\rad$; thus \mdot values from MHC98 and Gullbring et al. 1998, based on DM tracks, are modified to Lyon masses and radii, while \mdot from WG01, based on Lyon tracks, are adopted unchanged.  \mdot derived through $\hal$ modeling are rather insensitive to $\mass$ and $\rad$, so \mdot from MHCBH03 are also adopted unchanged.  (4) The WG01 sample comprises a number of binaries in Taurus.  Since they resolve the binaries with $HST$, and calculate \mdot and $\mass$ separately for each binary component, we are justified in including their measurements here (unlike in the MHC98 and WB03 cases mentioned in the last section, where we exclude the binaries because the components were not resolved, creating possible errors in the derived \mdot).  However, for a number of sources, they infer spectral types, and thereby masses, only from photometry.  Such types and masses are prone to error; we have excluded these sources from our analysis.  (5) We have {\it not} included here the study by N04, which cites \mdot inferred from near-IR spectroscopy.  The reason is that, for their $\rho$ Oph sources, spectral types and other (sub-)stellar parameters have been derived through IR spectral modeling; we cannot compare these quantitatively to our objects, which have values determined through completely different (optical) methodologies.  However, most of the sources examined by N04 (all their ChaI targets, and two of their $\rho$ Oph ones) have been independently observed by us in the optical, and are thus already present in our sample; excluding N04's IR study does not therefore significantly impoverish our sample size.  

The final \mdot--$\mass$ relationship is plotted in Fig. 6.  For all our low-mass objects, we plot the \mdot (or upper limits) computed via equation [3].   It is immediately clear that the accretion rate decreases sharply with mass.  A linear fit to the plot, {\it excluding} all points with only upper limits in accretion rate, yields \mdot $\propto$ ${\mass}^{2}$, all the way from higher-mass stars to the lowest mass BDs.  Using equation [1], instead of [3], to compute \mdot in the low-mass sample produces a nearly identical correlation (not plotted), with \mdot $\propto$ ${\mass}^{1.85}$.  The small difference in the two relationships is not significant: as Fig. 6 shows, there is considerable scatter in the plot, both in the high- and low-mass regimes, which makes any attempt to precisely pin down the \mdot--$\mass$ relationship physically unjustifiable (at least with the information at hand).  Under the circumstances, \mdot $\propto$ ${\mass}^{2}$ appears the simpler approximation to adopt at present.  This is not to say that the dispersion in the data is necessarily spurious.  Stochastic errors in our masses are unlikely to explain the scatter: while we might expect systematic uncertainties in our masses (since they are based on theoretical evolutionary tracks), the random errors should be a factor of $<$2 (since, regardless of any systematic offset in our masses, a well-determined spectral type should still correspond, from general principles, to a narrow range in \teff and hence mass, over the relatively small age-span of our sample).  Similarly, in both our analysis and the other studies we have used, \mdot errors are estimated to be a factor of $\sim$5, while the scatter in Fig. 6 is much larger: \mdot varies by $\pm$1.5 orders of magnitude at fixed mass.  A significant part of the observed scatter is thus probably real, reflecting real variations in accretion rate for a given mass.  Nevertheless, in spite of this dispersion, Fig. 6 presents clear evidence of an overall trend of declining accretion rate with decreasing mass.   

A similar \mdot--$\mass$ correlation has recently been noted by other investigators as well.  WB03 find their data to be consistent with a slope of \mdot $\propto$ $\mass$; however, their study contains only a single object below the sub-stellar boundary with measured \mdot.  Their relationship holds only when BDs with \mdot upper limits in their study are used in the fit, and thus is not inconsistent with our steeper slope.  MHCBH03, who do have some BDs with measured accretion rates, find $\dot{M}$ $\propto$ ${\mass}^2$, in agreement with our results.  Similarly, though N04 do not cite an explicit algebraic relationship, their stated numbers ($\dot{M}$ decreasing from $\sim$10$^{-8}$ to 3$\times$10$^{-10}$ \msun{yr$^{-1}$}, going from masses around 0.3--1\msun to $<$0.1\msun) are also consistent with $\dot{M}$ $\propto$ ${\mass}^2$.  However, even the latter two studies do not extend very far into the sub-stellar regime.  In particular, the N04 investigation, which has been the most complete one prior to the present work, and includes their own results from infrared spectroscopy as well as data from WB03, MHCBH03 and various other studies, still contains only $\sim$10 BDs ($\geq$M6.5 on the Luhman scale) with accretion signatures, all but one of which are M7 or earlier.  Our present sample, which includes 7 {\it new} accreting BDs (and 2 more possibly accreting ones), all but one of which range from M7 to M9, not only confirms the N04 and MHCBH03's mass--accretion rate relationship, but shows that the correlation extends unchanged over the entire sub-stellar domain, down to nearly the deuterium-burning limit.  

The physical basis for this correlation, however, remains unclear.  One obvious possibility is that the disk ionization decreases as one moves to cooler, lower-mass objects that produce less ionizing flux.  If the magneto-rotational instability is the underlying source of disk viscosity driving mass accretion, as seems to be the gathering consensus, then a declining disk ionization may be expected to produce lower \mdot.  This qualitative argument suggests that a fall-off in accretion rate with mass is perhaps not too surprising.  However, it does {\it not} explain why the fall-off should assume the particular functional form $\dot{M}$ $\propto$ ${\mass}^2$.  

Alternatively, Padoan et al. (2004) have recently suggested that the accretion rate is fundamentally controlled by Bondi-Hoyle accretion, wherein a forming star+disk system gathers mass as it moves through the surrounding medium.  The Bondi-Hoyle infall rate is:
$$ \dot{M}_{BH}  = \frac{4\pi\rho_{\infty}}{({c^2}_{\infty} + {v^2}_{\infty})^{3/2}}{\mathcal{M}}^2 \eqno[4]$$
where $\rho_{\infty}$, $c_{\infty}$ and $v_{\infty}$ are the gas density, sound speed and velocity of the surrounding material relative to the star-disk system, at large distances from the system.  $\mathcal{M}$ is formally the mass of the star+disk; for a CTT-like object, this is almost entirely the mass of the central star (or BD), so that $\mathcal{M} \sim \mass$.  {\it If} $\dot{M}_{BH}$ were equal to the disk accretion rate \mdot, then equation [4] would immediately imply $\dot{M}$ $\propto$ ${\mass}^2$, as indicated by the observations.  In reality, though, $\dot{M}_{BH}$ is the rate at which surrounding material lands on the star+disk system.  To explain the data, therefore, this scenario implicitly assumes a steady-state wherein the mass flux through the disk balances the mass flux falling on it at all times.  Whether this can always be achieved remains to be seen (e.g., one might imagine instead a situation where excess mass builds up in the disk, and is then removed by episodic increases in the disk accretion rate; \mdot would then equal $\dot{M}_{BH}$ only in the time-averaged sense, not at all times).   

Padoan et al. also point out that if their hypothesis is correct, then an intrinsic scatter can be expected as well in the \mdot--$\mass$ relationship: this would arise from random variations in the stellar velocities, gas densities and sound speeds, all of which affect \mdot by equation [4].  Padoan et al. then ascribe the empirical scatter seen by N04 in the \mdot--$\mass$ correlation, and reproduced in our Fig. 6, to such variations in physical conditions.  Their numerical simulations indicate resulting \mdot fluctuations of $\sim$ 1 dex, roughly consistent with the $\sim$ $\pm$1.5 dex envelope seen in our sample (Fig. 6).  However, this explanation of the observed scatter appears problematic, at least for the objects examined to date.  In both our and N04's investigation of \mdot versus of mass, the vast majority of the higher-mass CTTs (and many of the accreting VLMS and BDs as well) reside in Taurus; observationally, these objects still appear lined-up along the dense gas filaments in which they were presumable born (Hartmann 2002; Brice\~{n}o et al. 2002).  To be sure, these sources are unlikely to still be buried within the filaments, since their extinctions are generally relatively low; nevertheless, they do not appear to have moved very far from their birthplace.  Naively, then, one would expect their surrounding physical conditions to vary much less than over the whole cloud.  However, these sources still exhibit $\sim$3 orders of magnitude scatter in accretion rate at a given mass.  This is hard to understand in Padoan et al.'s scenario: indeed, to explain an \mdot dispersion of this magnitude, the latter authors invoke varying conditions over a large cloud volume certainly not restricted to the vicinity of the natal filaments.  Finally, denser surrounding material, and hence larger extinctions, might be expected towards stronger accretors in Padoan et al.'s scheme.  This is not apparent in the Taurus sample, however.  At a given mass, some of our CTTs differ by orders of magnitude in \mdot while exhibiting similar $\av$; similarly, most of the accreting Taurus BDs have $\av$ $\sim$ 0 while some non-accreting BDs in the same region have much higher extinctions.  In summary, the relevance of Bondi-Hoyle accretion to the observed \mdot--$\mass$ relationship is not yet completely clear.   

\subsection{Accretion and Age}
In Pre-Main Sequence stars, the accretion rate gradually falls off with age as the disk evolves and disk material is depleted (Hartmann et al. 1998).  Consequently, though WTTs - young stars with no measurable accretion - are found even in very young star-forming regions (presumably reflecting a spread in disk masses and accretion rates at any given age, plus a spread in ages within a given region), the fraction of CTTs - stars with clear signatures of disk accretion - clearly declines as one moves to significantly older regions.  For example, while the CTTs fraction in the $\sim$1.5 Myr old Taurus region is $\sim$50\% (Kenyon \& Hartmann 1995), it is only $\sim$15\% in the $\sim$10 Myr-old TW Hydrae Association (Muzerolle et al. 2000), and $\sim$1\% in the $\sim$15 Myr-old UCL+LLC sub-groups of the Sco-Cen Association (Mamajek, Meyer \& Liebert 2002).  It is of compelling interest to ask whether a similar decline in accretor fraction is also apparent in the VLMS and BD regime.  

Previous studies of young low-mass objects suggest that such a decline does exist.  JMB03 find that the proportion of VLMS and BDs with discernible accretion in high-resolution spectra decreases from the younger IC348 and Taurus star-forming regions to the older Upper Sco association.  MHCBH03 find a comparable drop as well, from IC348 and Taurus on the one hand to Upper Sco and $\sigma$ Ori on the other.  Through a combination of infrared and optical spectroscopy, N04 come to a similar conclusion, finding that the accretor fraction in the very young $\rho$ Oph region significantly exceeds that in the somewhat older ChaI cluster.  Finally, applying their $\hal$ equivalent-width criterion to low-resolution spectra to distinguish between accretion and activity, Barrado y Navascues \& Martin (2003; hereafter BM03) also find a steady decline in the proportion of low-mass accretors in going from younger (Taurus, IC348, ChaI) to older (Upper Sco, $\sigma$ Ori, TWHyA) regions.   

With the largest sample of VLMS and BDs observed with high-resolution spectroscopy to date, we can address this issue in more detail.  In particular, we wish to: {\it (1)} derive the accretor fraction in VLMS and BDs, and identify any signs of a decline with age; and {\it (2)} compare the accretor fraction in this low-mass sample to that in higher-mass T Tauri stars, to examine whether or not the accretor fractions in a given cluster, and the trend with age, are similar in the two mass ranges.  The spectral type ranges used to define ``low-mass objects'' and ``higher-mass T Tauri stars'' for this analysis are the same as employed throughout the paper: $\geq$M5 ($\lesssim$ 0.15 \msun) for the former, and K0--M4 (2--0.25 \msun) for the latter.  Moreover, of the young clusters and star-forming regions included in this paper, we will restrict our attention here to $\rho$ Oph, Taurus, IC348, ChaI and Upper Sco; only a total of 4 objects at $\geq$M5 are currently known in the TW HyA and R CrA regions, too few for any meaningful statistics.  

Our analysis proceeds as follows.  We first derive accretor fractions in the low-mass sample, by combining the results of our detailed accretion analysis (\S 5.1) with those of other similar recent studies.  This is described in {\it (i)} below, and provides our best estimate of the frequency of low-mass accretors in the examined regions.  However, these numbers cannot be compared directly to accretor fractions in higher-mass TTs, since most of the latter have not yet been rigorously analysed for the presence of accretion.  Instead, we estimate accretor fractions in the TTs using the simple BM03 criteria mentioned earlier (\S 5.1), and compare these to accretor fractions in the low-mass sample {\it re}-derived using the same criteria.  These calculations are described in {\it (ii)} and {\it (iii)} below.  While the latter estimates are less precise than those in {\it (i)}, they do allow a valid comparison between the low and higher-mass regimes.

{\it (i) Low-mass ($\geq$ M5) accretor fractions, from detailed accretion analysis}: All our high-resolution spectroscopy sources (Table 1) are included in this $\geq$ M5 sample, except MHO-9 (spectral type M4.25).  MHCBH03 and N04 have also conducted detailed optical and NIR studies of accretion; from these, we select 9 M5 and later sources in Taurus and IC348 (MHCBH03: CIDA-14, FN Tau, MHO-7, MHO-8, V410 Anon 13, V410 X-ray 5a, IC348 173, 407, 454) and 8 in $\rho$ Oph (N04: ISO-Oph 23, 32, 33, 102, 160, 176, 193, GY 10) that we have not observed\footnote{We have {\it not} included N04 sources in our previous \mdot vs. $\mass$ analysis, since N04 derive $\mass$ and \mdot through NIR calibrations of \teff, luminosity and extinction, which are not directly comparable to our optically calibrated values without analysis beyond the scope of this paper.  However, in the present analysis, we are interested only in whether the spectral types of the N04 sources are $\geq$M5 or not, and whether they are accreting ot not.  Through NIR comparisons, N04 find all their $\rho$ Oph sources to be $\geq$M6.  In the present study, we use optical spectral types; Luhman \& Rieke 1999 have shown that optical types for young objects are generally $\sim$1 sub-type earlier than NIR types for classes M6--M7, and $\sim$1--2 earlier for classes $\geq$M8.  Using this conversion, the N04 $\rho$ Oph sources would still have (optical) spectral types $\geq$M5; this expectation is borne out by comparing N04's types to those of Luhman \& Rieke for 4 sources in common.  Moreover, the qualitative identification of accretion by N04 in their sources is independent of the precise spectral type.  Thus, their sources can be included in our present analysis of accretor fraction at spectral types $\geq$M5.}.  GY 310 and IC348 256, which we designate as ``probably'' accreting in \S 5.1, are counted here as accretors.  However, we conservatively consider the ``possible'' accretors KPNO-Tau 4, KPNO-Tau 11 and CFHT-BD Tau 4, as well as the possible spectroscopic binaries KPNO-Tau 14 and DENIS 161929, as non-accretors in the present analysis.  For the MHCBH03 and N04 sources, we adopt the accretor / non-accretor status found by those authors.  Finally, since IC348 and ChaI appear to be at roughly the same age of $\sim$2 Myr, we do not differentiate between them in this accretion-age analysis, but instead combine their members into one group.  This serves three purposes: it provides a single large sample at $\sim$2 Myr, dilutes our bias towards accretors in the M5--5.5 range in IC348 (see \S2), and offsets the relatively small number of BDs (type $\geq$ M6) observed in ChaI alone.  Accretor fractions are calculated as $N_{acc}$/$N_{tot}$, and errors as $\sqrt{N_{acc}}$/$N_{tot}$.  

The resulting accretor fractions are: 58$\pm$22\% (7/12) in $\rho$ Oph, 32$\pm$10\% (10/31) in Taurus, 32$\pm$10\% (10/31) in IC348 + ChaI, and 5$\pm$5\% (1/20) in Upper Sco.  These are plotted in Fig. 7 (black triangles).  We see a distinct decrease in the frequency of accretion, moving from $\rho$ Oph at $\lesssim$1 Myr, to Taurus, IC348 and ChaI at 1.5--2 Myr, to Upper Sco at 5 Myr.  Given the error bars, the difference between $\rho$ Oph on the one hand and Taurus, IC348 and ChaI on the other is not definitive: comparable accretor fractions in these regions is not very surprising, since they are all close in age at $\lesssim$2 Myr.  However, a large fall-off in accretor frequency by Upper Sco ages is clear: accretion appears to have generally halted by $\sim$5 Myr (or more accurately, dropped below measurable levels by this age, corresponding to \mdot $\lesssim$ 10$^{-12}$ \msun{yr$^{-1}$} given our $\hal$ 10\% width $\gtrsim$ 200 \kms criterion for identifying acretors).   

{\it (ii) Low-mass ($\geq$ M5) accretor fractions, from BM03 criteria}: BM03 define spectral type-dependent $\hal$ equivalent width criteria to identify accretion.  We apply this technique to the same low-mass sample as above, to re-derive accretor fraction estimates that can be compared to T Tauri stars.  However, no $\hal$ measurements are available for 6 of N04's $\rho$ Oph sources included in our above sample; these are dropped from the present analysis.  Moreover, the BM03 criteria are defined for, and thus best applied to, low-resolution spectra.  We have thus used $\hal$ low-resolution equivalent widths collected from the literature for most of our low-mass sources\footnote{$\rho Oph$: Luhman \& Rieke 1999 (GY 141).  $Taurus$: Brice\~{n}o et al. 2002 \& Luhman et al. 2003b (KPNO sources); Mart\'{i}n et al. 2001 (CFHT sources); Brice\~{n}o et al. 1999 (CIDA-14); White et al. 1999 (GG Tau Ba,b); Brice\~{n}o et al. 1998 (MHO sources, V410 Anon 13, V410 X-ray 5a); Kenyon et al. 1998 (CIDA-1, FN Tau); Strom \& Strom 1994 (V410 X-ray 3).  $IC348$: Luhman et al. 2003a.  $ChaI$: Luhman 2004.  $Upper Sco$: Ardila, Mart\'{i}n \& Basri 2000 (USco sources); Mart\'{i}n, Delfosse \& Geieu 2004 (DENIS sources).}.  For a handful of objects, no low-resolution $\hal$ data are available; for these, we were forced to apply the BM03 scheme to high-resolution $\hal$ equivalent widths (GY 5, 37, 310 and USco 130, 131 from Table 1; ISO-Oph 32 and 102 from N04).  

The BM03 method implies low-mass accretor fractions of: 67$\pm$33\% (4/6) in $\rho$ Oph, 45$\pm$12\% (14/31) in Taurus, 52$\pm$13\% (16/31) in IC348 + ChaI, and 20$\pm$10\% (4/20) in Upper Sco.  These values are plotted in Fig. 7 (red circles), and are systematically higher than those inferred in {\it (i)} from a detailed study of accretion.  Compared to our high-resolution analysis, the BM03 scheme is more prone to mis-classifying high activity as accretion; conversely, it may be better at identifying high-inclination, low-\mdot accretors, whose high-resolution $\hal$ profiles would suggest activity alone (as discussed in \S 5.1).  Both effects yield a less conservative (higher) estimate of accretor fraction from BM03 criteria than derived in {\it (i)}.  Nevertheless, the two sets of estimates are comparable within the errors, and the BM03 fractions reveal the same trend with age: roughly constant accretor frequencies in low-mass objects from $\lesssim$1 to 2 Myr, followed by a large drop by the 5 Myr age of Upper Sco.    

{\it (iii) Higher-Mass T Tauri (K0--M4) accretor fractions, from BM03 criteria}: Finally, we use the BM03 technique to derive accretor fractions for higher-mass T Tauri stars, for comparison to the low-mass sample above.  BM03 themselves list accretor frequencies for T Tauri stars in various regions.  However, we prefer not to use their values, for three important reasons.  First, their `T Tauri' sample extends down to spectral type M5.5, which overlaps with our low-mass sample of $\geq$M5; as such, it does not provide a clean comparison between objects near the substellar boundary and those that are significantly higher in mass.  Second, their $\rho$ Oph T Tauri sample includes sources not just in the dark core of this region (where our low-mass $\rho$ Oph sources reside), but also in the surrounding areas.  Mart\'{i}n et al. (1998) have shown that the ratio of CTTs to WTTs decreases sharply outside the dark core; indeed, this probably accounts for the puzzlingly low CTTs fraction that BM03 find, and draw attention to, in $\rho$ Oph as a whole.  For a valid comparison between the low and high-mass regimes, therefore, we must select spatially proximate objects in $\rho$ Oph.  Finally, the BM03 `Upper Sco' sample includes sources in the Sco-Cen UCL+LLC sub-groups as well; since the latter are expected to be significantly older than the Upper Sco region alone (Mamajek, Meyer \& Liebert 2002), they are unsuitable for inclusion in our age analysis here.  

We have therefore compiled anew a list of K0--M4 stars in the pertinent star-forming regions and young clusters.  In $\rho$ Oph, all our low-mass objects lie within a $\sim$35$\arcmin$$\times$35$\arcmin$ area of the dark core originally studied by Casanova et al. 1995.  We select K0--M4 stars in this region from Bouvier \& Appenzeller 1992 and Mart\'{i}n et al. 1998.  For Taurus, we select stars from Brice\~{n}o et al. 2002 and Luhman et al. 2003b (since most of our Taurus low-mass objects lie in the Taurus fields analysed in these two studies).  For IC348 and ChaI, we select stars from Luhman et al. 2003a and Luhman 2004 respectively; the latter studies provide the currently most comprehensive list of stars in these regions.  Finally, for Upper Sco we select stars from the studies by Walter et al. 1994, Preibisch et al. 1998 and Prebisch et al. 2002, which together provide the most complete list of stars with available $\hal$ data in this cluster to date.  We have then compiled $\hal$ equivalent widths for all these stars from various sources in the literature\footnote{$\rho Oph$: Bouvier \& Appenzeller 1992; Mart\'{i}n et al. 1998.  $Taurus$: Hartmann et al. 1991; Hartigan, Strom \& Strom 1994; Mart\'{i}n et al. 1994; Mathieu 1994; Kenyon et al. 1998; Monin, M\'{e}nard \& Duch\^{e}ne 1998; Brice\~{n}o et al. 1998; Brice\~{n}o et al. 1999; White \& Hillenbrand 2004.  $IC348$: Luhman et al. 2003a.  $ChaI$: Luhman 2004.  $Upper Sco$: Walter et al. 1994; Preibisch et al. 1998; Preibisch et al. 2002.} -- mostly low-resolution, but high-resolution in a handful of cases where low-resolution data is unavailable -- and applied the BM03 accretion criteria.      

The resulting K0--M4 T Tauri accretor fractions (i.e., CTTs fractions) are:  50$\pm$16\% (10/20) in $\rho$ Oph; 59$\pm$9\% (42/71) in Taurus; 33$\pm$6\% (29/87) in IC348 and 44$\pm$8\% (28/63) in ChaI (so 38$\pm$5\% in IC348 + ChaI); and 7$\pm$2\% (12/170) in Upper Sco.  These are plotted in Fig. 7 (blue squares), and immediately point to two results.  

First, the decline in CTTs fraction is similar to that in the low-mass sample: while the CTTs frequencies are comparable in $\rho$ Oph, Taurus, IC348 and ChaI, they fall off steeply by the age of Upper Sco.  A closer look reveals that our CTTs fraction in ChaI+IC348, while consistent with that in $\rho$ Oph, is in fact somewhat lower than in Taurus.  The significance of this is unclear with the present errors; better number statistics, especially in $\rho$ Oph, can tell us whether there is a decline by the age of IC348 and ChaI, compared to the younger $\rho$ Oph and Taurus, or whether Taurus has a larger CTTs fraction relative to both the younger $\rho$ Oph and older IC48+ChaI regions.  At any rate, the data at hand show very clearly that the CTTs fraction has declined vastly by the age of Upper Sco, in agreement with the trend observed in the low-mass sample.  Second, in any given region, the high- and low-mass regimes evince comparable accretor fractions, within the errors (using BM03 criteria for both)\footnote{In Upper Sco, there appears to be a slight difference in accretor frequency between the two mass regimes, using the BM03 scheme, but the difference is not significant with our present low-mass number statistics.}.  

In summary, VLMS and BDs appear quite similar to higher mass T Tauri stars, both in the frequency of accretion at a given age, and in the overall accretion timescale (of order $\sim$5 Myrs).  It is worth noting that a few CTTs are also known in the TW Hydrae Association, which at 10 Myrs is older than Upper Sco.  We have not analysed TW HyA above, since only three low-mass objects have been identified in it so far (all BDs; Table 1); nevertheless, at least one of these sources is accreting as well (2MASS 1207-3932; see \S 5.1).  This supports our conclusion that accretion in low-mass objects can last just as long as in T Tauri stars.  

\subsection{Accretion and Disks}
Lastly, we examine the relationship between disk-accretion signatures, and infrared excesses directly indicating the presence of dusty disks.  Jayawardhana et al. (2003; hereafter JASH03) have carried out an $L'$-band survey of various young clusters and star-forming regions; they (conservatively) infer the presence of a disk when the $K-L'$ excess is $\geq$ 0.2 mag. After making a couple of minor revisions to their results (to account for updated spectral types and membership status for a few objects), we find that the frequency of disks in M5 and later objects is 67$\pm$27\% (6/9) in $\rho$ Oph, 55$\pm$25\% (5/9) in Taurus, 53$\pm$17\% (10/19) in IC348 + ChaI, and 60$\pm$35\% (3/5) in Upper Sco.  These values are consistent, within the errors, with a roughly constant disk fraction of 50--60\% from $\rho$ Oph down to Upper Sco ages.  This clearly differs from the trend in accretor fraction in the same regions, which, as we showed above, distinctly decreases by the 5-Myr age of Upper Sco.

To explore this issue further, we compare individual objects in common between JASH03's sample and ours.  First, the 6 accretors we have in common in $\rho$ Oph, Taurus, IC348 and ChaI all evince $K-L'$ excesses $>$ 0.2 mag.  Only one accretor in our study - 2MASS 1207-3932 in TWHyA - does not exhibit any $K-L'$ excess.  However, Apai et al. (2004) have recently detected disk-related mid-infrared emission from this source.  Their data suggest significant grain-evolution in the disk, which might account for the lack of an $L'$-band excess; alternatively, $L'$-band emission may be suppressed by a inner hole in the disk.  In any event, there is no evidence for accretion signatures arising in the complete absence of any disk excess, a situation that would be quite difficult to explain.  

However, there are a number of sources with $K-L'$ excess that do {\it not} have any measurable accretion (see Table 1).  For example, the ChaI sources ChaH$\alpha$ 1, 7, 8, 9 and 12, and the Upper Sco objects USco 104, 112 and 128, all have excesses greater than 0.2 mag, and are thus likely to possess disks; none of them appears to be accreting in our high-resolution spectra.  This strongly implies that significant amounts of disk material can persist around young VLMS and BDs, even after accretion has halted, or at least declined to unmeasurable levels.  This is consistent with our statistical argument above, showing that the disk fraction remains high even when the accretor fraction has diminished substantially with age.  This situation in the low-mass sample is analogous to the transition from the CTT to WTT phase in higher-mass stars: WTTs by definition do not have measurable accretion, yet often evince disk excesses (e.g., Haisch, Lada \& Lada 2001).  

\section{Concluding Remarks}
Our results may be summarized thus: {\it (1)} Classical T Tauri-like accretion is widespread in the very low-mass stellar and sub-stellar domain, extending to the least massive brown dwarfs examined so far; {\it (2)} accretion in these objects can be accompanied by mass outflows, as in CTTs; {\it (3)} the accretion rates decline sharply with mass, approximately as $\dot{M}$ $\propto$ ${\mass}^2$ (with considerable scatter); this functional form for the decline extends not only to the lowest mass stars, but also over the entire sub-stellar regime; {\it (4)} the accretor frequency decreases with age in very low-mass stars and brown dwarfs, as in higher-mass CTTs; {\it (5)} however, like in CTTs, accretion can continue in the low-mass regime for up to 10 Myrs; at any given age the accretor fraction in both mass regimes is comparable, implying similar accretion timescales; and {\it (6)} disks around objects near and below the hydrogen-burning limit persist beyond the main accretion phase, mirroring the transition from the classical to post-T Tauri phase in solar-type stars.  These phenomenological similarities suggest that very low-mass stars and brown dwarfs form in the same way as higher-mass stars do.  We find no evidence to support a change in formation mechanism between the two mass regimes, such as postulated in the `ejection' scenario.  

Nevertheless, it is worth remembering that the optical accretion diagnostics explored here, as well as the infrared excess measurements that attest to the ubiquity of sub-stellar disks, probe only the innermost disk radii.  It is still possible that outer disks are indeed truncated through dynamical ejections, while the inner regions survive for a long time merely by virtue of the small accretion rates in the low-mass regime.  Estimates of disk sizes (and masses) in brown dwarfs are required for a more conclusive resolution of this question.  The detection of optically thin sub-millimeter and millimeter emission, which would indicate the existence of cool outer disks, and also provide a direct estimate of dust (and, by extrapolation, disk) masses, would be vastly helpful in this regard.  Such emission has already been detected in two young brown dwarfs (Klein et al. 2003); larger surveys are currently underway, and should greatly clarify this issue of origins in the near future.

\acknowledgments
We would like to acknowledge the great cultural significance of Mauna Kea for native Hawaiians, and express our gratitude for permission to observe from atop this mountain.  We would also like to express our gratitude to the staff of the Keck and Magellan Observatories, who have made possible, and successful, our observations over the last several years.  We would especially like to thank John Johnson for the Magellan spectral reductions, and Russel White for kindly supplying two of the spectra used in this paper. S.M. would also like to thank Kevin Luhman for supplying some of the evolutionary-track-dependent data used in this paper, and also for many illuminating discussions on all aspects of brown dwarf formation.  This work was supported in part by University of Toronto start-up funds to R.J. and NSF grant AST-0098468 to G.B.  S.M. would like to acknowledge the support of the SIM-YSO grant for his postdoctoral research.

\plotone{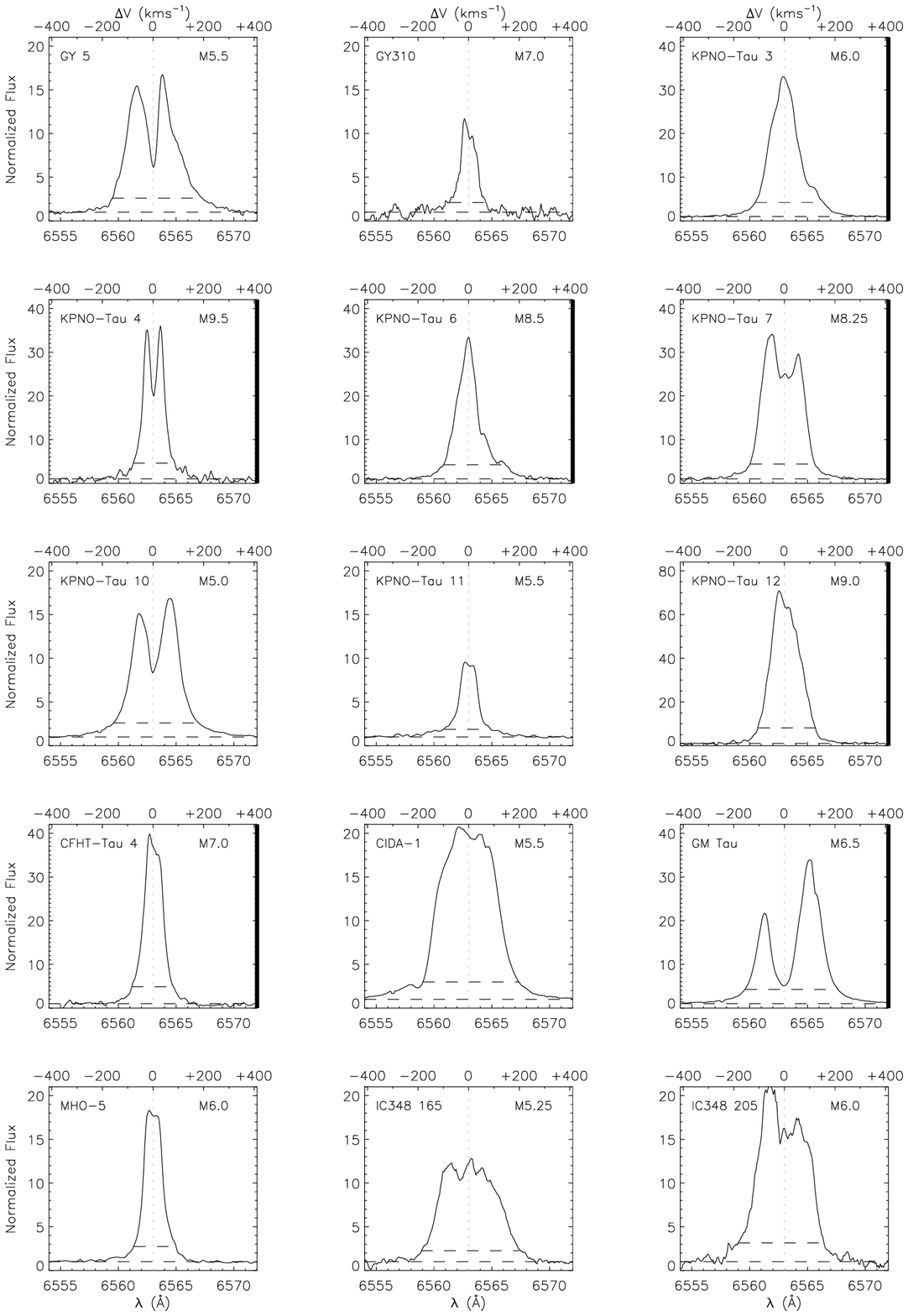}

\plotone{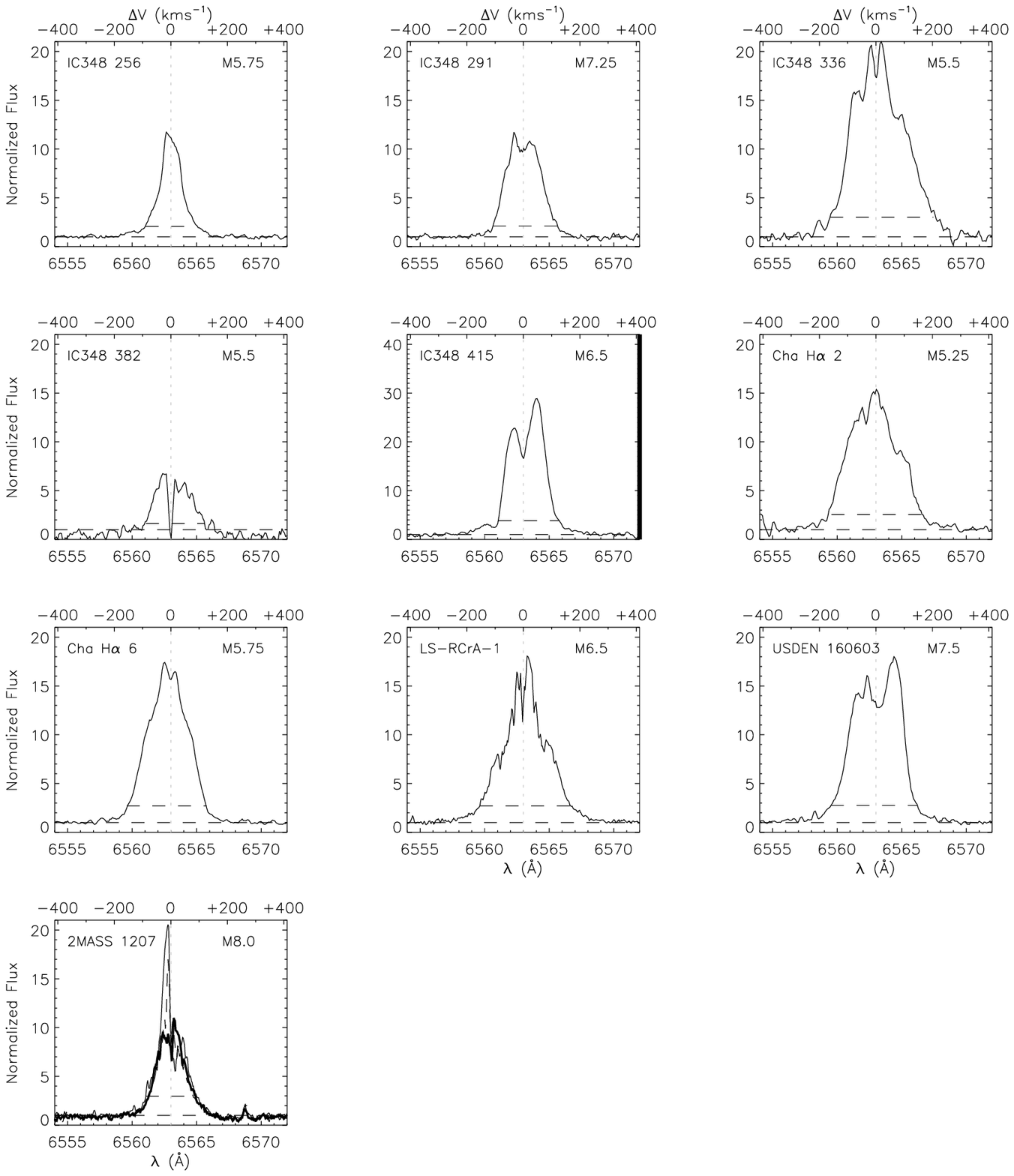}
\figcaption{\label{fig1} $\hal$ profiles for all our accretors, including the 5 objects designated ``probable'' and ``possible'' accretors in the text (\S5.1).  {\it Dashed} horizontal lines indicate the continuum and the $\hal$ 10\% width.  The emission is usually strong with broad line-wings, and often asymmetric.  Note that eight objects with very high $\hal$ equivalent widths are plotted on a y-axis scale larger than in the others; these panels are marked by a {\it thick} right-hand y-axis (in seven of these, the y-scale is twice that in the others; in KPNO-Tau 12, with the highest equivalent width in our sample, the scale is four times larger).  For LS-RCrA-1 (third panel from end), we have obtained 3 consecutive spectra, and show here the averaged $\hal$ profile (which does not vary much between the observations).  For 2MASS 1207-3932 (last panel), we also have 3 consecutive spectra; since its $\hal$ profile varies significantly over these $\sim$1 hour time-scales, we overplot all three (thick solid, thin solid, and dashed lines).  The separated profiles, and their discussion, can be found in Mohanty, Jayawardhana \& Navascu\'{e}s (2003).  }

\plotone{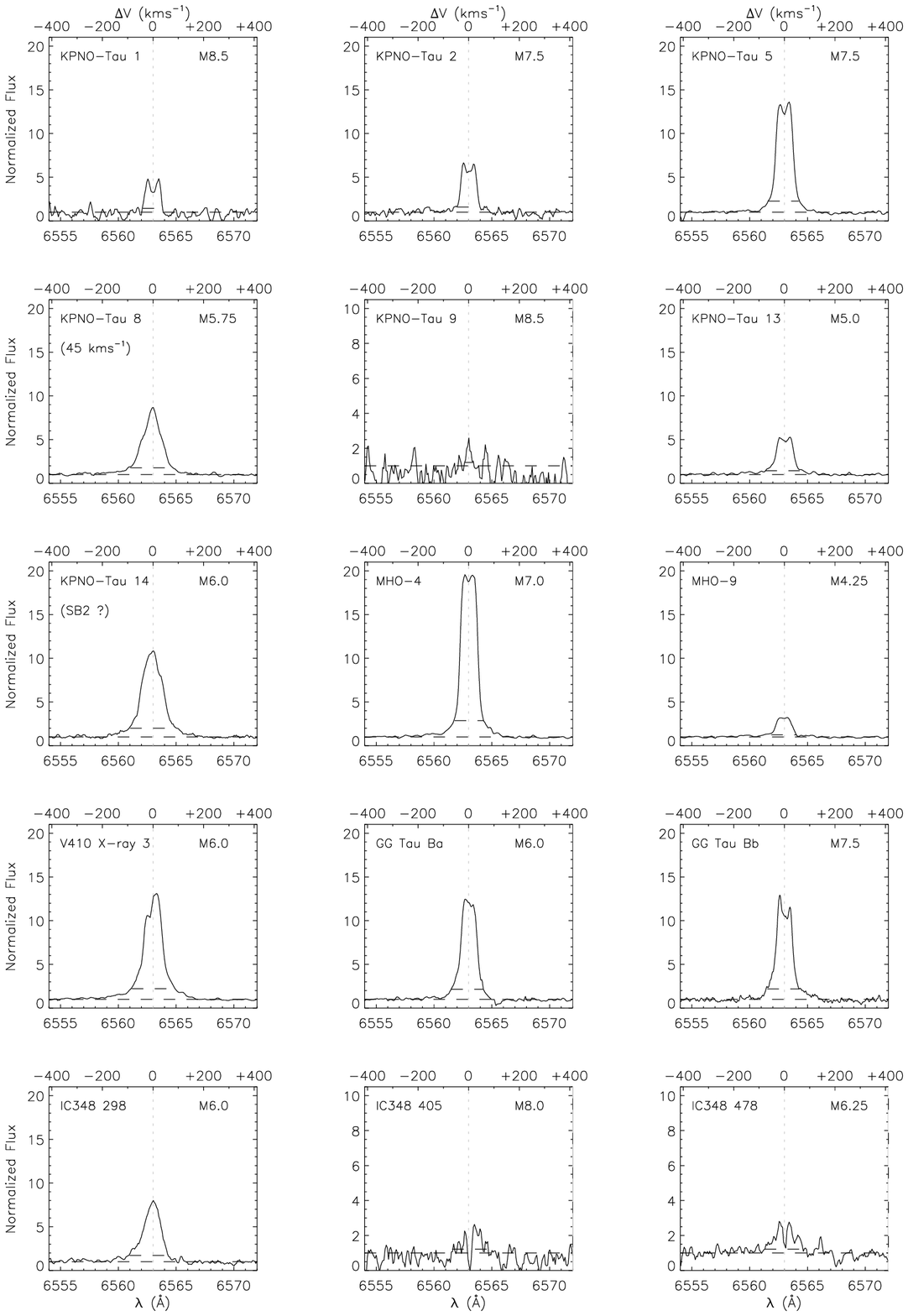}

\plotone{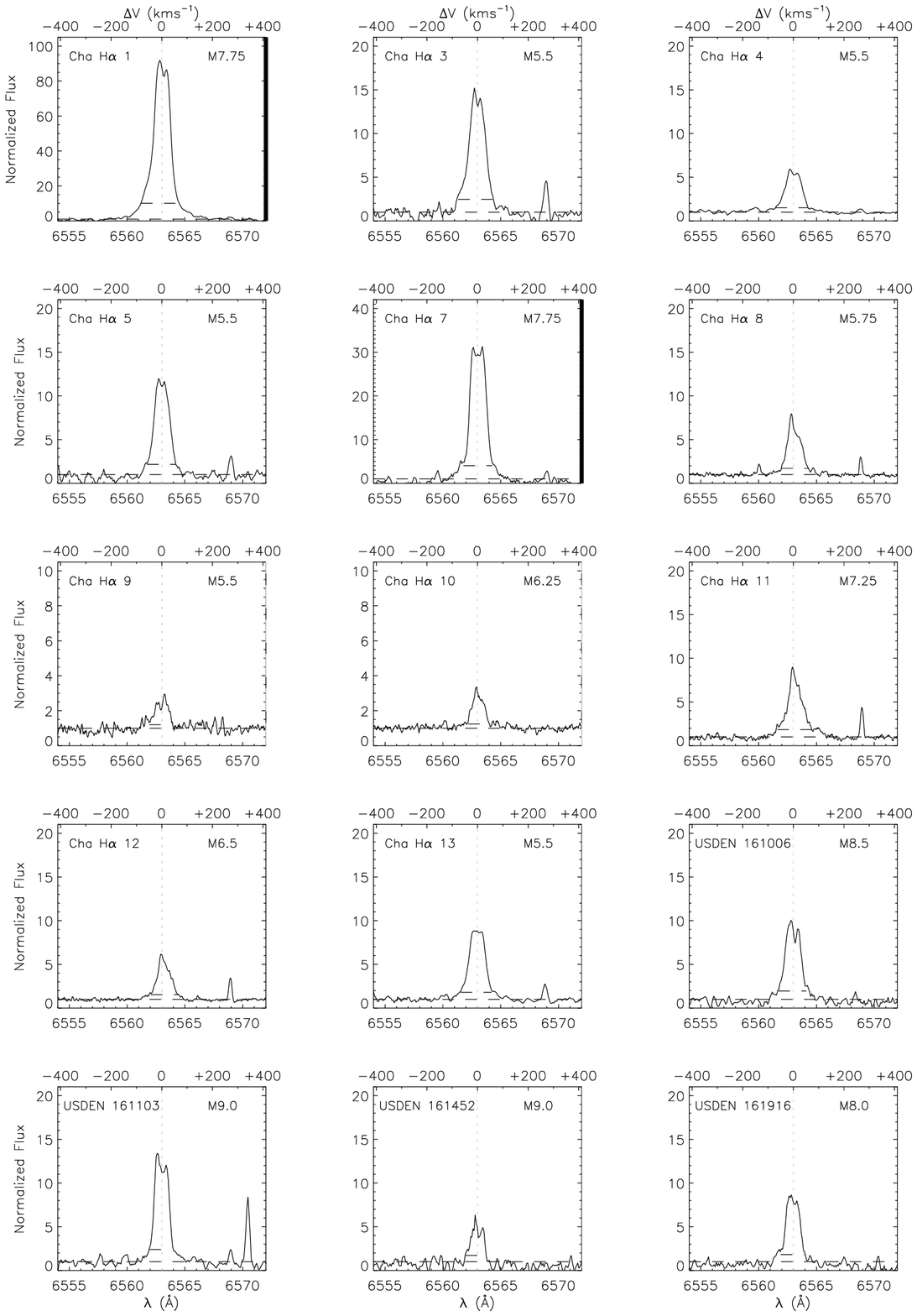}

\plotone{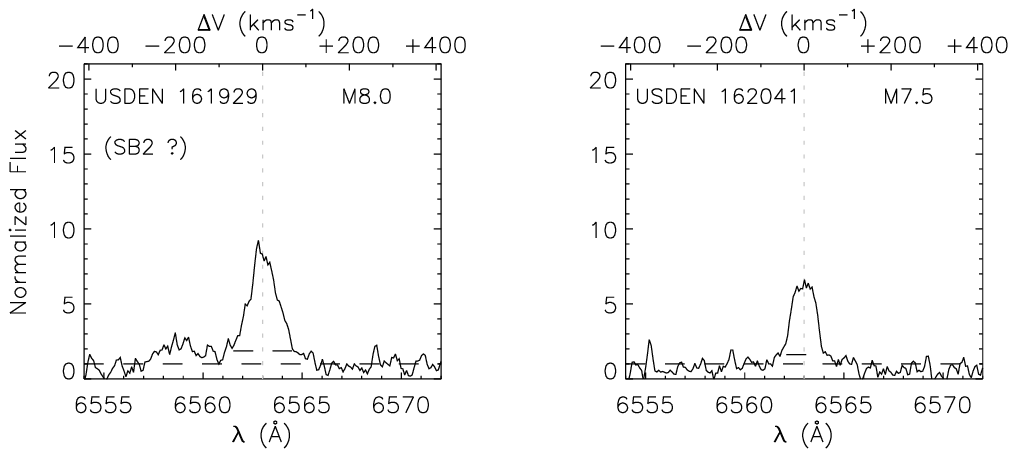}
\figcaption{\label{fig2} $\hal$ profiles for chromospherically active, {\it non}-accreting sources in our sample (only new objects are shown; the rest have been published previously -- see \S 5.1 and Table 1).  Compared to the accretors (Fig.1), the emission here is usually much weaker, narrower, and symmetric.  Four objects with extremely small emission are plotted on a y-axis scale half that in most others; these panels are marked with a {\it dashed} right-hand y-axis.  Two sources with very strong (but still narrow and symmetric) emission are plotted on a y-scale larger than the others: by a factor of five in Cha $\hal$ 1 and a factor of two in Cha $\hal$ 7 (panels marked with a {\it thick} right-hand y-axis).  The $\hal$ profile in KPNO-Tau 8 is comparatively broad, due to its high \vsini (45 \kms); KPNO-Tau 14 and USco DENIS 161929 appear to be spectroscopic binaries from our cross-correlation analysis, probably accounting for the somewhat broad profile in the former and asymmetry in the latter.  Emission is also asymmetric in Cha $\hal$ 11 and 12; they might be accreting, but do not evince any other signatures of doing so in our data (\S5.1), so we classify them as active for now.  Note that the emission at $\sim$6569\AA~ in some panels is a sky-line, not stellar, that remained unsubtracted.  }

\plotone{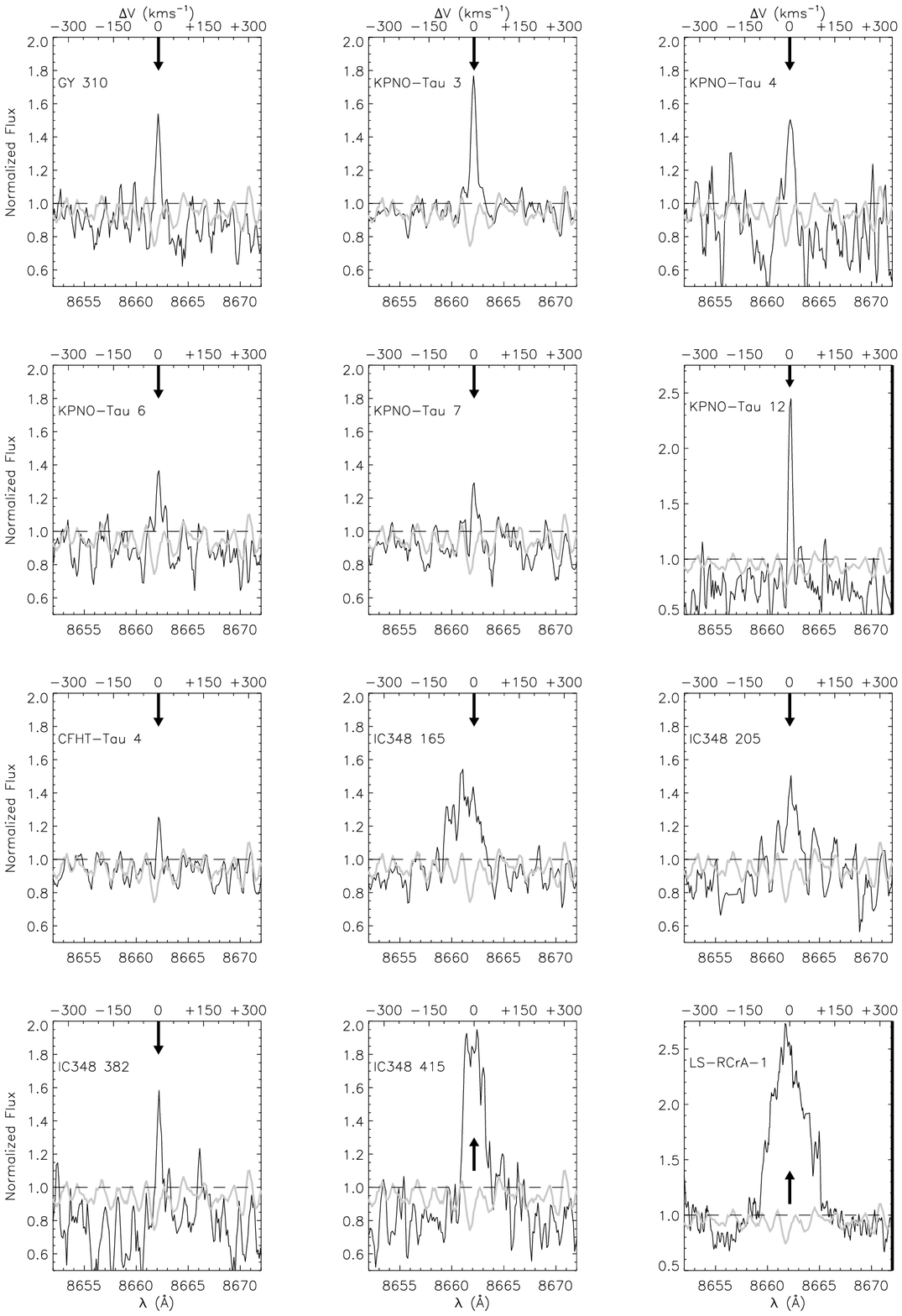}
\figcaption{\label{fig3}  Observed \cal \lam8662.14 emission in our accretors.  The emitting objects are plotted in {\it black}; the emission line is marked with an arrow; the horizontal {\it dashed} line is the normalized continuum.  For comparison, we overplot in {\it grey} the same spectral region for KPNO-Tau-5, a young, non-accreting M7.5 object in Taurus with good S/N and no \cal emission.  The comparison shows that most of the bumps and wiggles in our accretors, in the continuum around the \cal line, are real (there are some mismatches due to varying spectral type -- our accretors range from M5 to M9.5 -- and low S/N in a few sources).  In particular, the excellent agreement between KPNO-Tau 5 and our weakest emitters -- KPNO-Tau 6, KPNO-Tau 7 and CFHT-BD-Tau 4 -- at all points in the continuum except only at the \cal position, shows that \cal emission is indeed present in the latter objects.  Note that the dip in the spectrum of KPNO-Tau-5 at the approximate position of \cal is {\it not} due to photospheric absorption in this line: the photosphere is much too cool for this.  Instead, this dip is just another wiggle in the continuum that happens to overlap with the \cal position.  The observed emission in most of our accretors is fairly narrow, with FWHM $\lesssim$ 50\kms.  Nevertheless, as described in \S5.1, the emission is seen only in the accretors; in merely chromospherically active low-mass sources, the spectrum simply looks like that of KPNO-Tau 5, which is active as well (evincing chromospheric $\hal$).  }

\plotone{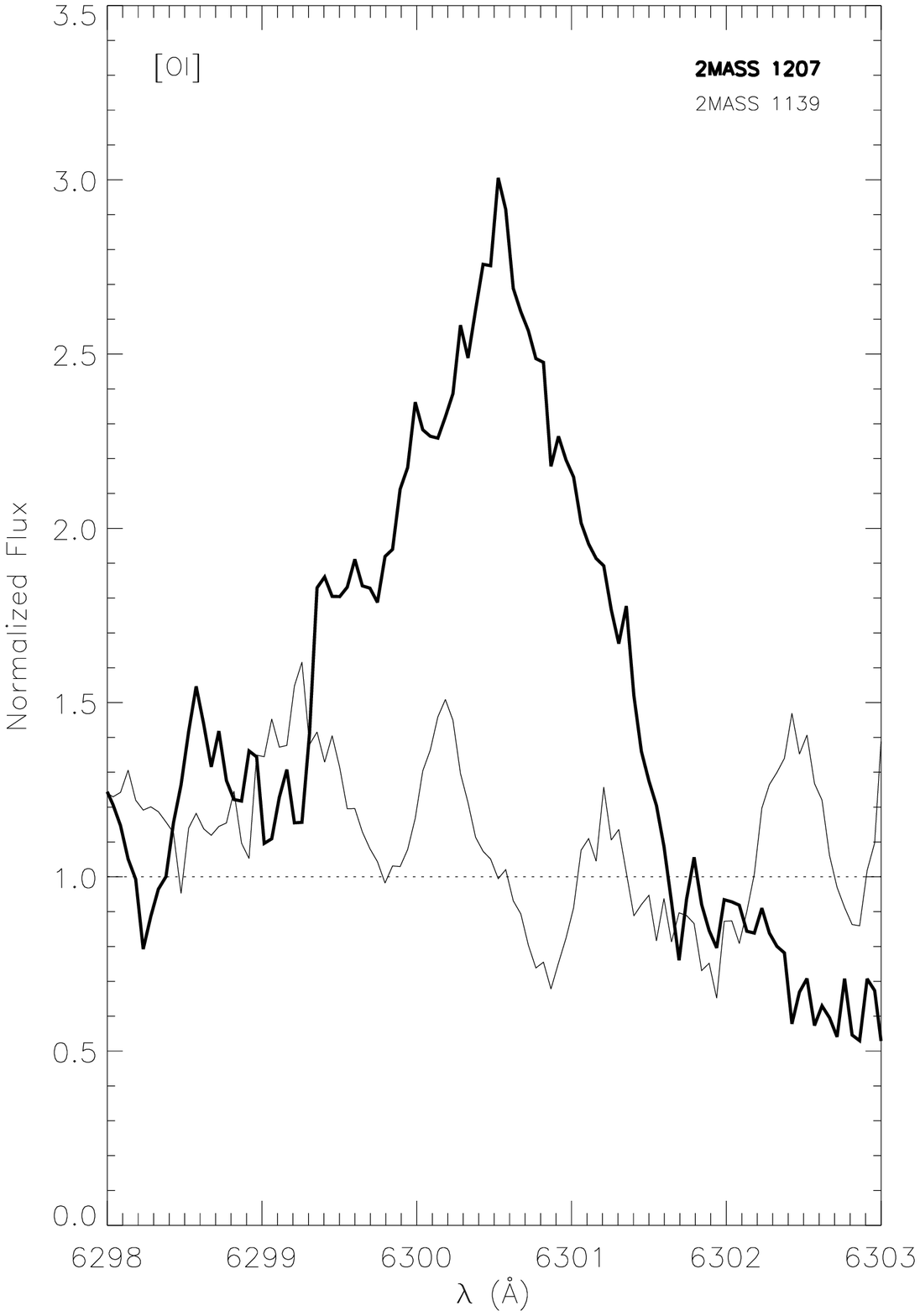}
\figcaption{\label{fig4} Forbidden [\oxy] \lam6300 emission in the accreting M8 TW HyA brown dwarf 2MASS 1207-3932 ({\it thick} line).  The horizontal dashed line is the continuum (normalized over a much larger wavelength range than plotted).  For comparison, we overplot the spectrum of 2MASS 1139-3159 ({\it thin} line), which is a non-accretor with the same spectral type, also in TW HyA.  Excess emission is clearly visible in the accretor, indicating that mass outflows can persist, like accretion, in low-mass BDs for up to 10 Myr.}

\plotone{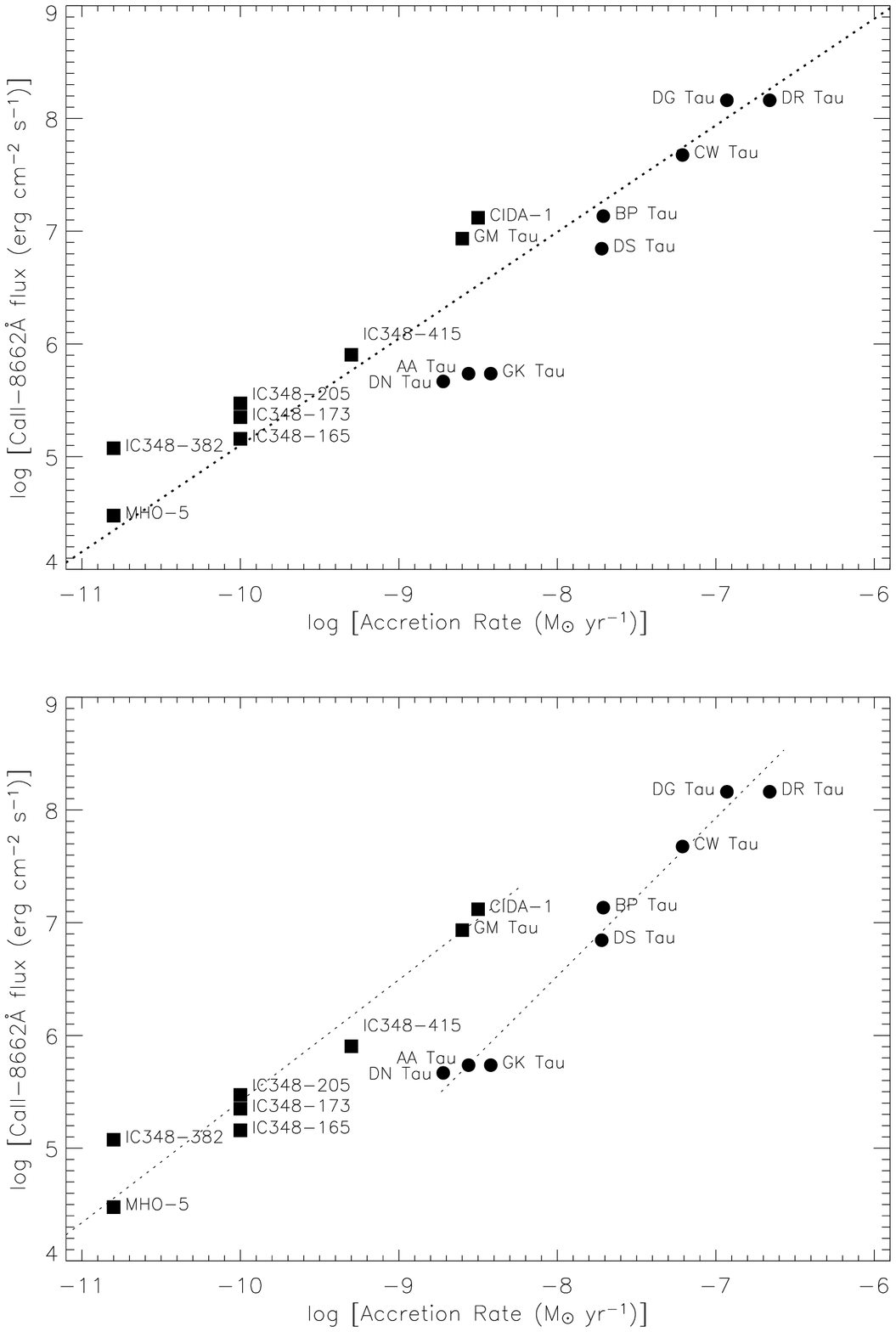}
\figcaption{\label{fig5} Observed dependence of \cal emission-flux on \mdot (where \mdot is calculated independently of the \cal flux, from either veiling or $\hal$-modeling; \S5.2).  Higher-mass CTTs are shown as {\it filled circles}, and low-mass accretors (VLMS and BDs) as {\it filled squares}.  The \cal flux declines sharply with decreasing accretion rate over the entire sample, with the correlation extending over $\sim$4 orders of magnitude in both flux and \mdot.  {\it Top panel}:  Single fit to both the CTTs and low-mass accretors (equation [1], \S 5.2).  Notice that the low-mass sample lies slightly above the overall fit, while most of the CTTs lie slightly below.  {\it Bottom panel}:  Separate fits to the CTTs and low-mass accretors (equations [2] and [3], \S 5.2).  The sytematic offset between the CTTs and low-mass objects is now obvious, with the CTTs evincing lower \cal flux at a given accretion rate.  The scatter around the individual fits is much reduced, compared to the single fit in the top panel.  See text (\S 5.2). }



\plotone{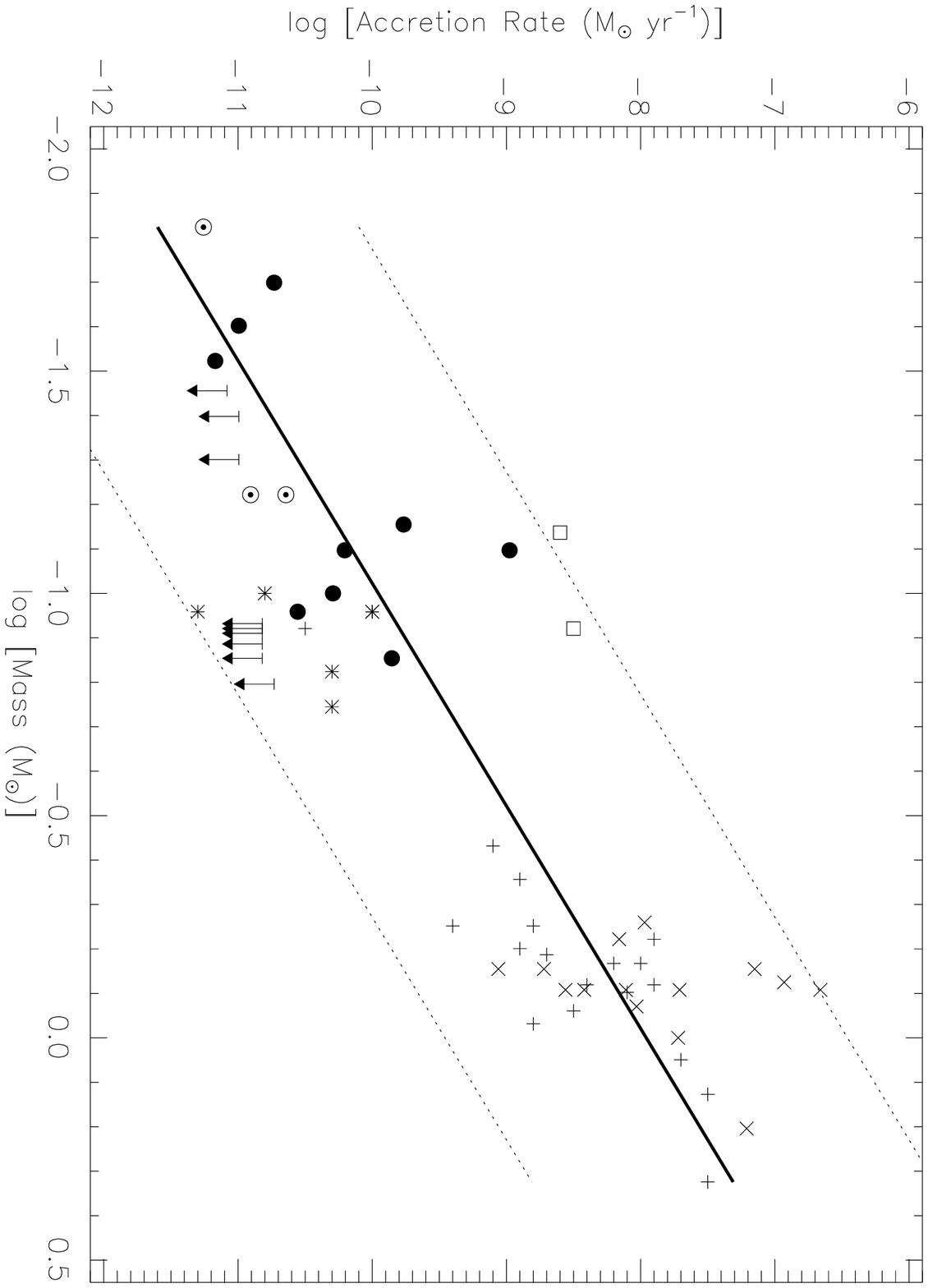}
\figcaption{\label{fig6} Derived relationship between \mdot and mass.  Low-mass accretors with \mdot derived in this paper, from \cal fluxes using eqn.[3], are shown as either {\it filled circles} (bona-fide accretors with detected \cal), {\it bulls-eyes} (``probable'' or ``possible'' accretors with detectd \cal) or {\it downward arrows} (accretors without detected \cal emission: \mdot upper limits derived from \cal detection limit of 0.1\AA).  {\it Crosses} are CTTs from Gullbring et al. 1998; {\it pluses} are CTTs in binaries from WG01; {\it squares} are low-mass sources from WB03; {\it asterisks} are low-mass sources from MHCBH03 (see \S5.3).  The thick {\it solid black} line is our formal fit to the data (excluding the upper limit points): $\dot{M}$ $\propto$ ${\mass}^{2}$.  The two {\it dashed} lines are this fit vertically offset by $\pm$1.5 dex, to denote the upper and lower envelopes of the trend.  All sources lie within (or very close to) this envelope, suggesting that the data is consistent with scatter around a single functional relationship between \mdot and $\mass$ such as we have derived (however, the data is not sufficient to prove this).  The main implication is that \mdot decreases sharply with mass, following the same slope from CTTs to VLMS to the lowest mass BDs: over nearly 2.5 orders of magnitude in mass and 4 orders in \mdot.  See text (\S 5.3). }

\plotone{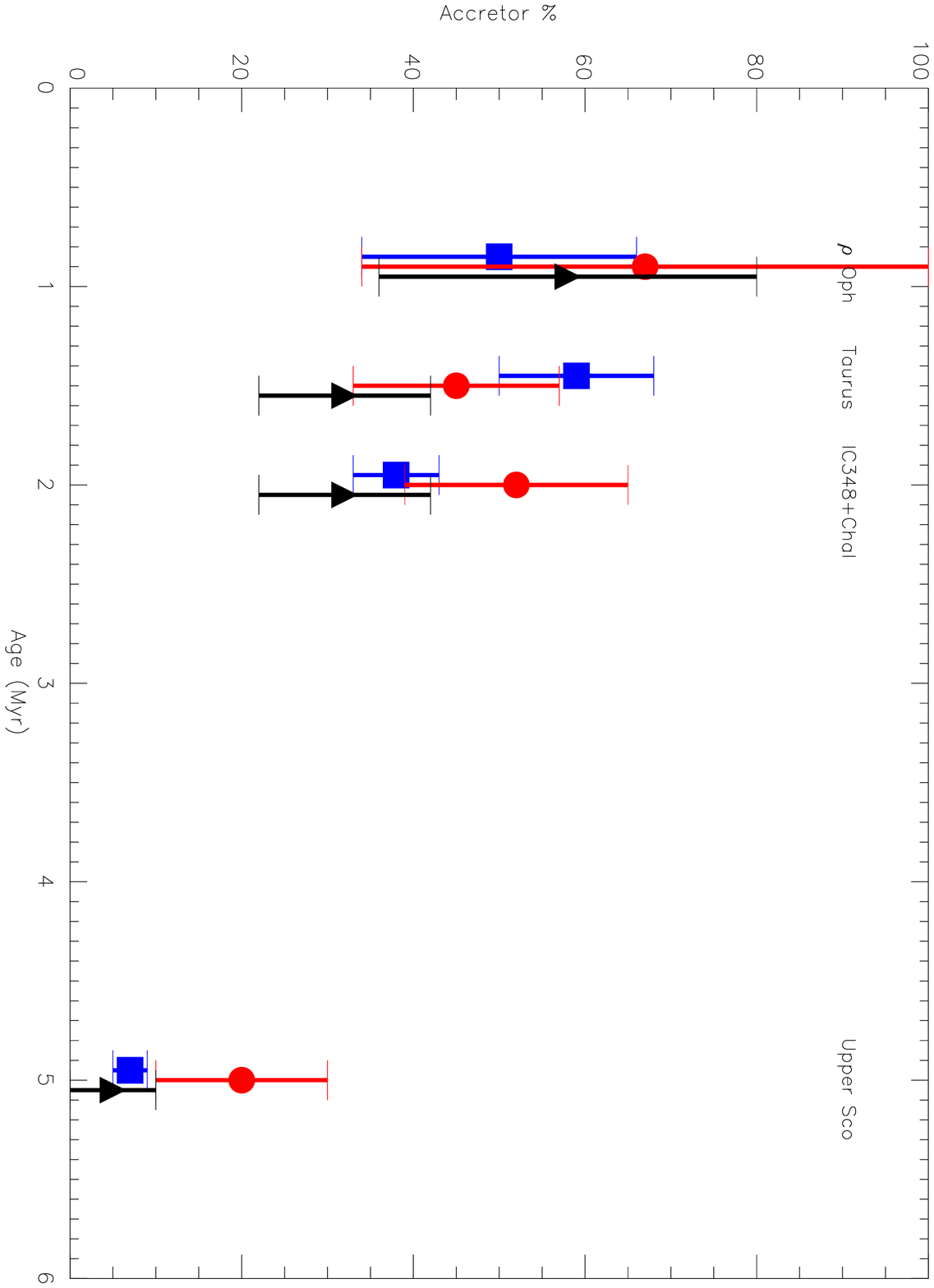}
\figcaption{\label{fig7} Accretor fraction (\%) as a function of age.  The regions selected to represent various ages are marked: $\rho$ Oph ($\lesssim$1 Myr), Taurus (1.5 Myr), IC348+ChaI (2 Myr; members of these two coeval regions are grouped together), and Upper Sco (5 Myr).  {\it Black triangles}: accretor fractions in low-mass sources (spectral type $\geq$M5), from a detailed analysis of accretion signatures.  {\it Red circles}: accretor fractions in the low-mass sources, now using BM03 criteria instead.  {\it Blue squares}: accretor fractions in higher mass stars (K0--M4), again using BM03 criteria.  At a given age, the three symbols are slightly offset horizontally for clarity.  We see that (1) accretor fractions from BM03 criteria are systematically higher than from detailed analysis, in the low-mass sample; (2) the accretor fractions in both the low-mass sample (using either BM03 criteria or detailed analysis) and the higher-mass stars fall sharply by 5 Myr; and (3) at any fixed age, the accretor fractions in the low- and high-mass regimes are comparable (using BM03 criteria for both).  See text (\S 5.4).  }  

\clearpage
\begin{landscape}
\begin{deluxetable}{llcccccclccc}
\tablecaption{\label{tab1}Rotation, Emission Lines and Accretor Status in Our Sample}
\tablewidth{0pt}
\tabletypesize{\scriptsize}
\tablehead{
\colhead{name} &
\colhead{SpT}  &
\colhead{\vsini} &
\colhead{$\hal$} &
\colhead{$\hal$ 10\% width}  &
\colhead{HeI \lam6678}  &
\colhead{OI \lam8446}  &
\colhead{CaII \lam8662} &
\colhead{accretor?\tablenotemark{a}} &
\colhead{comment\tablenotemark{b}} &
\colhead{region} &\\
 & &[\kms] &[\AA] &[\kms] &[\AA] &[\AA] & [\AA] &$\quad$[Y/N] & & \\}   
                            
\startdata

GY 5              & M5.5  & 16.0  & 64.9  & 352.  & 0.6    & --      & --     & Y  & disk  & $\rho$ Oph \\
GY 37             & M6.0  & 22.5  & 5.1   & 102.  & --     & --      & --     & N  & --    & $\rho$ Oph \\
GY 141            & M8.5  & 6.0   & 13.4  & 87.   & --     & --      & --     & N  & --    & $\rho$ Oph \\
GY 310            & M7.0  & 10.0  & 17.2  & 144.  & --     & --      & 0.2    & Y (probable)  & disk  & $\rho$ Oph \\
KPNO-Tau 1        & M8.5  & 5.5   & 4.6   & 77.   & --     & --      & --     & N  & disk  & Taurus \\
KPNO-Tau 2        & M7.5  & 0.0   & 7.7   & 88.   & --     & --      & --     & N  & --    & Taurus \\
KPNO-Tau 3        & M6.0  & 0.0   & 97.7  & 269.  & 1.6    & 0.1     & 0.5    & Y  & disk  & Taurus \\
KPNO-Tau 4        & M9.5  & 10.0  & 68.4  & 155.  & --     & --      & 0.3    & Y (possible)  & --    & Taurus \\
KPNO-Tau 5        & M7.5  & 10.0  & 21.8  & 121.  & --     & --      & --     & N  & --    & Taurus \\
KPNO-Tau 6        & M8.5  & 5.0   & 77.5  & 244.  & 1.6    & 0.2     & 0.2    & Y  & disk  & Taurus \\
KPNO-Tau 7        & M8.25 & 0.0   & 122.  & 255.  & 0.9:   & --      & 0.1    & Y  & disk  & Taurus \\
KPNO-Tau 8        & M5.75 & 45.0  & 15.0  & 167.  & --     & --      & --     & N  & --    & Taurus \\
KPNO-Tau 9        & M8.5 & --    & 0.7:  & 32:    & --     & --      & --     & N  & disk    & Taurus \\
KPNO-Tau 10       & M5.0  & 18.0  & 66.7  & 336.  & 0.4    & 0.5     & --     & Y  & --    & Taurus \\
KPNO-Tau 11       & M5.5  & 11.0  & 17.7  & 182.  & 0.3    & --      & --     & Y (possible) & --    & Taurus \\
KPNO-Tau 12       & M9.0  & 5.0   & 207.  & 224.  & 1.1:   & 1.1     & 0.6    & Y  & --    & Taurus \\ 
KPNO-Tau 13       & M5.0  & 5.0   & 7.5   & 135.  & --     & --      & --     & N  & --    & Taurus \\
KPNO-Tau 14       & M6.0  & --    & 22.1  & 170.  & --     & --      & --     & N  & SB2?  & Taurus \\
CFHT-BD-Tau 1     & M7.0  & 7.0   & 7.4   & 85.   & --     & --      & --     & N  & --    & Taurus \\
CFHT-BD-Tau 2     & M7.5  & 8.0   & 7.2   & 76.   & --\tablenotemark{c}& --   & -- & N  & --    & Taurus \\
CFHT-BD-Tau 3     & M7.75 & 12.0  & 43.   & 129.  & --     & --      & --     & N  & --    & Taurus \\
CFHT-BD-Tau 4     & M7.0  & 11.0  & 79.   & 164.  & --\tablenotemark{c} & --  & 0.1    & Y (possible)  & disk  & Taurus \\
GG Tau Ba         & M6.0  & 8.0   & 19.3  & 125.  & --     & --      & --     & N  & --    & Taurus \\
GG Tau Bb         & M7.5  & 7.0   & 19.0  & 125.  & --     & --      & --     & N  & --    & Taurus \\
CIDA-1            & M5.5  & 5.0   & 112.  & 378.  & 0.9:   & Y       & 23.6   & Y  & --    & Taurus \\
GM Tau            & M6.5  & 9.0   & 120.  & 365.  & 0.2    & Y       & 14.5   & Y  & --    & Taurus \\
MHO-4             & M7.0  & 7.0   & 33.   & 115.  & 0.1    & --      & --     & N  & --    & Taurus \\
MHO-5             & M6.0  & 8.0   & 37.   & 166.  & --     & --      & --     & N(Y)\tablenotemark{d}&--    & Taurus \\
MHO-9             & M4.25 & 10.0  & 3.4   & 99.   & --     & --      & --     & N  & --    & Taurus \\
V410 Xray 3       & M6.0  & 14.0  & 27.   & 173.  & --     & --      & --     & N  & --    & Taurus \\
IC348 165         & M5.25 & 19.0  & 66.   & 389.  & 0.7    & 1.1:    & 1.1    & Y  & disk  & IC348 \\
IC348 205         & M6.0  & 6.0   & 93.   & 338.  & --     & 1.9     & 0.4    & Y  & --    & IC348 \\
IC348 256         & M5.75 & 9.0   & 23.   & 180.  & 0.4    & --      & --     & Y (probable)  & disk  & IC348 \\
IC348 286         & M5.75 & 19.0  & 6.9   & 148.  & --     & --      & --     & N  & --    & IC348 \\
IC348 291         & M7.25 & 0.0   & 32.4  & 228.  & --     & 1.1     & --     & Y  & --    & IC348 \\
IC348 298         & M6.0  & 22.5  & 12.7  & 148.  & --     & --      & --     & N  & --    & IC348 \\
IC348 336         & M5.5  & 11.0  & 89.6  & 365.  & 2.2:   &         & --     & Y  & --    & IC348 \\
IC348 353         & M6.0  & 25.0  & 5.8   & 117.  & --     & --      & --     & N  & --    & IC348 \\
IC348 355         & M8.0  & 45.0  & 10.3  & 235.  & --     & --      & --     & N  & --    & IC348 \\
IC348 363         & M8.0  & 14.0  & 3.8   & 82.   & --     & --      & --     & N  & --    & IC348 \\
IC348 367         & M5.75 & 20.0  & 6.8   & 92.   & --     & --      & --     & N  & disk  & IC348 \\
IC348 382         & M5.5  & 10.0  & 15.   & 208.  & --     & 1:      & 0.2     & Y  & --    & IC348 \\
IC348 405         & M8.0  & 45:   & 1.7:   & 141: & --     & --      & --      & N  & --    & IC348 \\
IC348 415         & M6.5  & 0.0   & 80.   & 213.  & --     & 3.4     & 1.7     & Y  & --    & IC348 \\
IC348 478         & M6.25 & 20.0  & 3.1   & 150.  & --     & --      & --     & N  & --    & IC348 \\
Cha H$\alpha$ 1   & M7.75 & 8.0   & 173.  & 151.  & --     & --      & --     & N  & disk  & Cha I \\
Cha H$\alpha$ 2   & M5.25 & 20.0  & 63.3  & 317.  & --     & --      & --     & Y  & disk  & Cha I \\
Cha H$\alpha$ 3   & M5.5  & 22.0  & 24.0  & 136.  & --     & --      & --     & N  & --    & Cha I \\
Cha H$\alpha$ 4   & M5.5  & 20.0  & 8.3   & 127.  & --     & --      & --     & N  & --    & Cha I \\
Cha H$\alpha$ 5   & M5.5  & 16.0  & 17.8  & 111.  & --     & --      & --     & N  & --    & Cha I \\
Cha H$\alpha$ 6   & M5.75 & 5.0   & 59.6  & 282.  & --     & --      & --     & Y  & disk  & Cha I \\
Cha H$\alpha$ 7   & M7.75 & 11.0  & 52.2  & 114.  & --     & --      & --     & N  & disk  & Cha I \\
Cha H$\alpha$ 8   & M5.75 & 10.0  & 8.4   & 106.  & --     & --      & --     & N  & --    & Cha I \\
Cha H$\alpha$ 9   & M5.5  & 9.0   & 2.4   & 91.   & --     & --      & --     & N  & disk  & Cha I \\
Cha H$\alpha$ 10  & M6.25 & 10.0  & 2.5   & 83.   & --     & --      & --     & N  & --    & Cha I \\
Cha H$\alpha$ 11  & M7.25 & 16.0  & 12.2  & 136.  & --     & --      & --     & N  & --    & Cha I \\
Cha H$\alpha$ 12  & M6.5  & 22.5  & 6.8   & 107.  & --     & --      & --     & N  & disk  & Cha I \\
Cha H$\alpha$ 13  & M5.5  & 14.0  & 14.3  & 122.  & --     & --      & --     & N  & --    & Cha I \\
LS-RCrA-1         & M6.5  & 18.0  & 53.   & 316.  & --     & Y       & 9.6    & Y  & --    & R CrA \\
USco 40           & M5.0  & 37.5  & 7.9   & 137.  & --     & --      & --     & N  & --    & Upper Sco \\
USco 53           & M5.0  & 45.0  & 17.8  & 175.  & --     & --      & --     & N  & --    & Upper Sco \\
USco 55           & M5.5  & 12.0  & 7.3   & 114.  & --     & --      & --     & N  & --    & Upper Sco \\
USco 66           & M6.0  & 27.5  & 6.5   & 115.  & --     & --      & --     & N  & --    & Upper Sco \\
USco 67           & M5.5  & 18.0  & 12.9  & 139.  & --     & --      & --     & N  & --    & Upper Sco \\
USco 75           & M6.0  & 63.0  & 8.9   & 212.  & --     & --      & --     & N  & --    & Upper Sco \\
USco 100          & M7.0  & 50.0  & 13.1  & 184.  & --     & --      & --     & N  & --    & Upper Sco \\
USco 104          & M5.0  & 16.0  & 9.4   & 109.  & --     & --      & --     & N  & disk  & Upper Sco \\
USco 109          & M6.0  & 6.0   & 6.3   & 84.   & --     & --      & --     & N  & --    & Upper Sco \\
USco 112          & M5.5  & 8.0   & 9.5   & 111.  & --     & --      & --     & N  & disk  & Upper Sco \\
USco 128          & M7.0  & 0.0   & 15.9  & 121.  & 0.3    & --      & --     & N  & disk  & Upper Sco \\
USco 130          & M7.0  & 14.0  & 8.4   & 111.  & --     & --      & --     & N  & --    & Upper Sco \\
USco 131          & M7.0  & 16.0  & 14.2  & 105.  & --     & --      & --     & N  & --    & Upper Sco \\
USco DENIS 160603 & M7.5  & 9.0   & 70.   & 306.  & --     & --      & --     & Y  & --    & Upper Sco \\
USco DENIS 161006 & M8.5  & 7.0   & 13.9  & 110.  & --     & --      & --     & N  & --    & Upper Sco \\
USco DENIS 161103 & M9.0  & 0.0   & 16.9  & 90.   & --     & --      & --     & N  & --    & Upper Sco \\
USco DENIS 161452 & M9.0  & 0.0   & 5.3   & 79.   & --     & --      & --     & N  & --    & Upper Sco \\
USco DENIS 161916 & M8.0  & 5.0   & 10.3  & 93.   & --     & --      & --     & N  & --    & Upper Sco \\
USco DENIS 161929 & M8.0  & --  & 13.6  & 135.    & --     & --      & --     & N  & SB2    & Upper Sco \\
USco DENIS 162041 & M7.5  & 16.0  & 7.4   & 81.   & --     & --      & --     & N  & --    & Upper Sco \\
2MASS 1207-3932   & M8.0  & 13.0  & 27.7  & 204.  & 0.8    & --      & --     & Y  & disk  & TW HyA \\
2MASS 1139-3159   & M8.0  & 25.0  & 7.3   & 111.  & --     & --      & --     & N  & --    & TW HyA \\  
TWA 5B            & M8.5  & 16.0  & 5.1   & 162.  & --     & --      & --     & N  & --    & TW HyA \\  

\enddata

\tablenotetext{a}{Accreting or not, as discussed \S5.1.  Non-accretors marked with `N', ``bon-fide'' accretors marked with `Y' (including MHO-5; see [d] below), and ``probable'' and ``possible'' accretors marked as such (see \S5.1).  }
\tablenotetext{b}{Objects with known direct evidence for disks, from excesses in the mid-IR (2MASS 1207-3932) and/or in the near-IR $L'$-band (all others).  Note that sources with no information may still have disks.}
\tablenotetext{c}{Our Keck spectra for CFHT-BD-Tau 2 and 4 do not cover the \hel \lam6678 line.  However, we do not see any overt evidence for \hel emission in their published low-resolution spectra (Mart\'{i}n et al. 2001). }
\tablenotetext{d}{MHO-5 looks like a non-accretor in our spectrum, but shows forbidden [\oxy] \lam6300, strongly indicative of an outflow, and hence accretion, in MHCMH03's data ([\oxy] is not covered in our spectrum); we thus assume it is a bona-fide accretor (see \S 5.1).}

\end{deluxetable}
\end{landscape}

\begin{deluxetable}{llccccclcccc}
\tablecaption{\label{tab2}Objects Used to Derive \cal Flux -- Accretion Rate Relationship}
\tablewidth{0pt}
\tabletypesize{\scriptsize}
\tablehead{
\colhead{name} &
\colhead{SpT}  &
\colhead{\teff}  &
\colhead{$\vcal$\tablenotemark{a}}  &
\colhead{CaII \lam8662} &
\colhead{$\fcal$\tablenotemark{b}}  &
\colhead{$\rad$\tablenotemark{c}}  &
\colhead{$\mass$\tablenotemark{c}}  &
\colhead{\mdot\tablenotemark{d}} &
\colhead{{acc}-method\tablenotemark{e}} &
\colhead{ref\tablenotemark{f}} &\\
 & &[K] & &[\AA] & [$log$(erg s$^{-1}$ cm$^{-2}$)]& [R$_{\odot}$]& [M$_{\odot}$] & [$log$(M$_{\odot}$ yr$^{-1}$)] &  & \\}

\startdata
AA Tau    & K7    & 4050 & 0.0  & 0.5   & 5.74 & 2.14 & 0.78  & -8.56 & ex     & 1 \\
BP Tau    & K7    & 4050 & 0.6  & 7.8   & 7.13 & 2.14 & 0.78  & -7.71 & ex     & 1 \\
CW Tau    & K3    & 4750 & 0.8  & 11.3  & 7.68 & 2.94 & 1.60  & -7.21 & ex     & 1 \\
DG Tau    & K7/M0 & 3950 & 3.6  & 30.3  & 8.16 & 2.13 & 0.75  & -6.93 & ex     & 1 \\ 
DN Tau    & M0    & 3850 & 0.0  & 0.5   & 5.67 & 2.10 & 0.70  & -8.72 & ex     & 1 \\
DR Tau    & K7    & 4050 & 5.1  & 21.8  & 8.16 & 2.14 & 0.78  & -6.66 & ex     & 1 \\
DS Tau    & K5    & 4350 & 0.3  & 3.3   & 6.84 & 2.32 & 1.00  & -7.72 & ex     & 1 \\
GK Tau    & K7    & 4050 & 0.0  & 0.5   & 5.74 & 2.14 & 0.78  & -8.42 & ex     & 1 \\
\\
CIDA-1    & M5.5  & 3050 & 0.69 & 23.6  & 7.12 & --   & 0.12  & -8.5  & ex     & 2 \\
GM Tau    & M6.5  & 2950 & 1.04 & 14.5  & 6.93 & --   & 0.073 & -8.6  & ex     & 2 \\
\\
MHO-5     & M6.0  & 3000 & 0.0  & 0.1   & 4.48 & --   & 0.10  & -10.8 & $\hal$ & 3 \\
IC348 165 & M5.25 & 3100 & 0.0  & 0.4   & 5.16 & --   & 0.14  & -10.0 & $\hal$ & 3 \\
IC348 173 & M5.75 & 3025 & 0.0  & 0.7   & 5.35 & --   & 0.11  & -10.0 & $\hal$ & 3 \\
IC348 205 & M6.0  & 3000 & 0.1  & 0.9   & 5.47 & --   & 0.10  & -10.0 & $\hal$ & 3 \\
IC348 382 & M5.5  & 3050 & 0.2  & 0.3   & 5.07 & --   & 0.11  & -10.8 & $\hal$ & 3 \\
IC348 415 & M6.5  & 2950 & 0.1  & 2.6   & 5.90 & --   & 0.07  & -9.3  & $\hal$ & 3 \\

\enddata

\tablenotetext{a}{Veiling in the vicinity of \cal \lam8662 (adopting the values calculated at 8600\AA~ for sources in ref. (1), 8400\AA~ for sources in ref. (2), and 8900\AA~ for sources in ref. (3)).}
\tablenotetext{b}{\cal \lam8662 emission flux, calculated by us by combining the observed equivalent width and veiling with the continuum flux estimated for the given spectral type (\S2).}
\tablenotetext{c}{Mass and radius from the Lyon tracks, for the given spectral types (see \S2).  Radii are calculated only for those sources where they are needed to modify the \mdot (see [d], [e], [f] below).  }
\tablenotetext{d}{Accretion rate from various sources, modified by us from the original values in some cases (\S5.1; see notes [e] and [f] below).  }
\tablenotetext{e}{Method used to derive \mdot.  `ex': from optical veiling excess; `$\hal$': from $\hal$ line-profile modeling. }
\tablenotetext{f}{References for the spectral types, veiling, \cal \lam8662 equivalent widths and \mdot.  {\it 1}: MHC98; {\it 2}: WB03; {\it 3}: MHCBH03.  MHC98 use DM tracks to infer \mdot from veiling, so we have modified their original \mdot to Lyon masses and radii (\S5.1); these modified \mdot (and new masses and radii) are cited here.  The MCHBH03 \mdot from $\hal$-modeling are adopted unchanged; we have only changed the masses they cite to Lyon values.  The WB03 values for \mdot (from UV excess) and mass (from Lyon tracks) are both adopted unchanged.  Finally, WB03 do not supply \cal \lam8662 equivalent widths for GM Tau and CIDA-1, but we have calculated these from the same spectra they used. }

\end{deluxetable}

\begin{deluxetable}{llccclllc}
\tablecaption{\label{tab3} Derived Masses and Accretion Rates in Our Sample}
\tablewidth{0pt}
\tabletypesize{\scriptsize}
\tablehead{
\colhead{name} &
\colhead{SpT}  &
\colhead{\teff}  &
\colhead{$\vcal$}  &
\colhead{$\fcal$}  &
\colhead{$\mass$}  &
\colhead{{\mdot}\tablenotemark{a}} &
\colhead{{\mdot}\tablenotemark{b}} & \\
 & &[K] & & [$log$(erg s$^{-1}$ cm$^{-2}$)]& [M$_{\odot}$] & [$log$(M$_{\odot}$ yr$^{-1}$)] & [$log$(M$_{\odot}$ yr$^{-1}$)] \\}

\startdata
GY 5          & M5.5  & 3050 & 0.0  & $<$4.52 & 0.12  & $<$-10.6 & $<$-10.8 (-10.8, -10.1)\tablenotemark{c}\\
GY 310        & M7.0  & 2900 & 0.0  & 4.70    & 0.06  & -10.4: & -10.6: (--- , -9.3)\tablenotemark{c}\\
KPNO-Tau 3    & M6.0  & 3000 & 0.0  & 5.18    & 0.08  & -9.9 & -10.2 \\
KPNO-Tau 4    & M9.5  & 2300 & 0.0  & 4.08    & 0.015 & -11.1: & -11.2: \\
KPNO-Tau 6    & M8.5  & 2550 & 0.0  & 4.34    & 0.025 & -10.8 & -11.0 \\ 
KPNO-Tau 7    & M8.25 & 2650 & 0.0  & 4.18    & 0.03  & -11.0 & -11.2 \\
KPNO-Tau 10   & M5.0  & 3125 & 0.0  & $<$4.58 & 0.16  & $<$-10.5 & $<$-10.7 \\
KPNO-Tau 11   & M5.5  & 3050 & 0.0  & $<$4.52 & 0.13  & $<$-10.6: & $<$-10.8: \\
KPNO-Tau 12   & M9.0  & 2400 & 0.0  & 4.62    & 0.02  & -10.5 & -10.7 \\
CFHT-BD-Tau 4 & M7.0  & 2900 & 0.0  & 4.40    & 0.06  & -10.7: & -10.9: \\
MHO-5         & M6.0  & 3000 & 0.0  & $<$4.48 & 0.10  & $<$-10.7 &  $<$-10.9 (-10.8)\tablenotemark{c} \\
IC348 165     & M5.25 & 3100 & 0.0  & 5.60    & 0.14  & -9.5 & -9.9 \\
IC348 205     & M6.0  & 3000 & 0.0  & 5.08    & 0.10  & -10.0 & -10.3 \\
IC348 256     & M5.75 & 3025 & 0.0  & $<$4.51 & 0.12  & $<$-10.6: & $<$-10.8: \\
IC348 291     & M7.25 & 2850 & 0.0  & $<$4.36 & 0.05  & $<$-10.8 & $<$-11.0 \\ 
IC348 382     & M5.5  & 3050 & 0.0  & 4.82    & 0.11  & -10.3 & -10.6 \\
IC348 415     & M6.5  & 2950 & 0.0  & 5.68    & 0.07  & -9.4 & -9.8 \\
Cha $\hal$ 2  & M5.25 & 3100 & 0.0  & $<$4.56 & 0.14  & $<$-10.6 & $<$-10.8 (-10.0, -10.8)\tablenotemark{c}\\    
Cha $\hal$ 6  & M5.75 & 3025 & 0.0  & $<$4.51 & 0.12  & $<$-10.6 & $<$-10.8 (-10.5, $<$-10.8)\tablenotemark{c}\\
LS-RCrA-1     & M6.5  & 2950 & 0.15 & 6.50    & 0.08  & -8.5 & -9.0 \\
USco DENIS 160603 & M7.5 & 2800 & 0.0 & $<$4.32 & 0.04  & $<$-10.8 & $<$-11.0 \\
2MASS 1207-3932 & M8.0 & 2700 & 0.0   & $<$4.20 & 0.035  & $<$-10.9 & $<$-11.1 \\
\\
CIDA-1        & M5.5  & 3050 & 0.69 & 7.12    & 0.12  & --- & --- (-8.5)\tablenotemark{d} \\
GM Tau        & M6.5  & 2950 & 1.04 & 6.93    & 0.073 & --- & --- (-8.6)\tablenotemark{d} \\
IC348 336     & M5.5  & 3050 & ---   & ---    & 0.10  & --- & --- (-10.0)\tablenotemark{d} \\

\enddata

\tablenotetext{a}{Accretion rates for our low-mass sample, inferred from the \cal \lam8662 fluxes via equation [1] (i.e., using a simultaneous fit to the higher-mass CTTs and low-mass accretors; \S 5.2).  For sources with no detected \cal emission, upper limits for \mdot are given, based on our detection limit of 0.1\AA~ for the \cal equivalent width.  For the 5 sources classified as ``probable'' or ``possible'' accretors (Table 1 and \S 5.1), our derived \mdot are marked with a colon. }
\tablenotetext{b}{Same as previous column, but now \mdot calculated using equation [3] (i.e., using the fit to only low-mass accretors; \S 5.2).}
\tablenotetext{c}{For objects in which we have derived \mdot upper limits, but \mdot measurements have been made by other investigators as well, we cite the latter values in parantheses in addition to our upper limits.  For MHO-5, we cite \mdot from MHCBH03 (from $\hal$ modeling); for GY 5, GY 310, Cha $\hal$ 5 and Cha $\hal$ 6, we cite \mdot from N04 (from both $\hal$ modeling and NIR Pa$\beta$ luminosities, in that order).  While MHCMH03's and N04's values are higher than our upper limits in some cases, we do not consider this a serious problem: the discrepancies are generally quite small, and within an order of magnitude in all but one case; they are within the range expected from the differences in our techniques for determining the (sub-)stellar parameters and \mdot, as well as from possible \mdot variations in time (indeed, similar differences are present even within N04's values, derived using two different techniques).  }
\tablenotetext{d}{For GM Tau, CIDA-1 and IC348 336, we do {\it not} calculate \mdot from \cal fluxes, but adopt instead the \mdot values from WB03 (GM Tau, CIDA-1) and MHCBH03 (IC348 336), cited in parantheses.  For GM Tau and CIDA-1, we have used our \cal fluxes, combined with the WB03 \mdot, to {\it derive} the \mdot--\cal relationships (see Table 2), so we cannot use these relationships to derive \mdot again from the same \cal fluxes.  In IC348 336, we do not see any \cal \lam8662 emission or any clear signs of veiling which might hide the line.  However, its TiO bands around 8440\AA~ do not match too well those of other sources at the same spectral type.  The mismatch does not seem to result from veiling, though MHCBH03 find a small possible veiling of $\sim$0.1$\pm$0.1 at 7100 and 8900\AA.  The TiO discrepancy may be due to uncertainty in its spectral type (within 1 subclass) which we have adopted from Luhman et al. 2003a.  As such, we choose not to derive an \mdot upper limit based on our \cal non-detection, but adopt instead the \mdot found by MHCBH03 from $\hal$ modeling. }

\end{deluxetable}

\begin{deluxetable}{lccccc}
\tablecaption{\label{tab4} Masses and Accretion Rates in Other CTTs and Low-Mass Accretors}
\tablewidth{0pt}
\tabletypesize{\scriptsize}
\tablehead{
\colhead{name} &
\colhead{SpT}  &
\colhead{$\mass$\tablenotemark{a}}  &
\colhead{\mdot\tablenotemark{b}} & 
\colhead{ref\tablenotemark{c}} & \\
 & & [M$_{\odot}$] & [$log$(M$_{\odot}$ yr$^{-1}$)] &  \\}

\startdata

CY Tau     & M1   & 0.60 & -8.16 & 1 \\
DE Tau     & M2   & 0.55 & -7.97 & 1 \\
DO Tau     & M0   & 0.70 & -7.15 & 1 \\
GI Tau     & K6   & 0.85 & -8.03 & 1 \\
GM Aur     & K7   & 0.78 & -8.11 & 1 \\
IP Tau     & M0   & 0.70 & -9.06 & 1 \\
\\
DD Tau A   & M3   & 0.37 & -9.1 & 2 \\
DF Tau     & M0.5 & 0.68 & -8.0  & 2 \\
FO Tau A   & M2   & 0.60 & -7.9  & 2 \\
FV Tau A   & K5   & 1.12 & -7.7 & 2 \\
GG Tau Aa  & K7   & 0.76 & -7.9  & 2 \\
----------- Ab & M0.5 & 0.68 & -8.2  & 2 \\
----------- Ba & M6.0 & 0.12 & -10.5 & 2 \\
GH Tau A   & M1.5 & 0.63 & -8.9  & 2 \\
----------- B  & M2  & 0.56 & -9.4  & 2 \\
IS Tau A   & K7   & 0.79 & -8.1  & 2 \\
RW Aur A   & K1   & 1.34 & -7.5 & 2 \\
----------- B  & K5 & 0.93 & -8.8 & 2 \\ 
T Tau A    & K0   & 2.11 & -7.5 & 2 \\
UZ Tau A   & M1   & 0.65 & -8.7  & 2 \\  
----------- Ba & M2   & 0.56 & -8.8  & 2 \\
V807 Tau A & K7   & 0.76 & -8.4  & 2 \\
V955 Tau A & K5   & 0.87 & -8.5  & 2 \\
XZ Tau A   & M3   & 0.44 & -8.9  & 2 \\
\\
CIDA-14      & M5    & 0.15 & -10.3 & 3 \\ 
MHO-6        & M4.75 & 0.18 & -10.3 & 3 \\ 
V410 Anon 13 & M5.75 & 0.11 & -11.3 & 3 \\ 

\enddata

\tablenotetext{a}{Masses from Lyon tracks.}
\tablenotetext{b}{Accretion rates from veiling, UV-excess and $\hal$ profile modeling, in references [1], [2] and [3] respectively (see note [c] below). }
\tablenotetext{c}{References for the spectral types and \mdot.  {\it 1}: Gullbring et al. 1998; {\it 2}: WG01; {\it 3}: MHCBH03.  The Gullbring et al. \mdot, calculated from $U$-band veiling using DM tracks, have been modified here to Lyon masses and radii; the WG01 \mdot, from UV-excess but using Lyon tracks, have been adopted unchanged. The MHCBH03 \mdot from $\hal$ modeling are also adopted unchanged, but the masses given in in MHCBH03, from DM tracks, have been changed to Lyon ones. See \S5.3.}

\end{deluxetable}

\end{document}